\documentclass[10pt]{article}

   \usepackage[utf8]{inputenc}
   \usepackage{graphicx}
   \usepackage{amsfonts}
   \usepackage{amssymb}
   \usepackage{amsthm}
   \usepackage{amsmath}
   \usepackage{bbold}
   \usepackage[active]{srcltx}

   \usepackage{here}
   \usepackage{cite}
   \usepackage{color}
   \usepackage{soul}

   \usepackage[affil-it]{authblk}

   \usepackage{xcolor}
   \usepackage[
   bookmarks=true, linktocpage, colorlinks=true, allbordercolors=white, citecolor=blue]{hyperref}
   \usepackage{xspace}

   \graphicspath{{figs/}}

\newcommand{\be}{\begin{equation}}
\newcommand{\ee}{\end{equation}}

\newcommand{\beq}{\begin{equation}}
\newcommand{\eeq}{\end{equation}}
\newcommand{\non}{\nonumber}
\newcommand\bea{\begin{eqnarray}}
\newcommand\eea{\end{eqnarray}}

\allowdisplaybreaks
\newcommand\tb{\tan\beta}

\newcommand\ReDiag{\mathop{%
  \raise .5pt\hbox{[}%
  \widetilde{\mathrm{Re}}%
  \raise .5pt\hbox{]}}}
\newcommand\ReOffDiag{\mathop{%
  \raise .5pt\hbox{$\llbracket$}%
  \widetilde{\mathrm{Re}}%
  \raise .5pt\hbox{$\rrbracket$}}}

\newcommand\Mh{M_h}

\newcommand\MA{M_A}

\newcommand\mb{m_b}
\newcommand\mt{m_t}

\newcommand\refeq[1]{Eq.~(\ref{#1})}
\newcommand\refeqs[1]{Eqs.~(\ref{#1})}
\newcommand\refta[1]{Table~\ref{#1}}
\newcommand\refse[1]{Sect.~\ref{#1}}

\newcommand\citere[1]{Ref.~\cite{#1}}
\newcommand\citeres[1]{Refs.~\cite{#1}}

\newcommand{\gev}{\,\, \mathrm{GeV}}

\newcommand{\br}{\text{BR}}

\newcommand{\sig}{\sigma}

\def\reffi#1{\mbox{Fig.~\ref{#1}}}

\def\ga{\gamma}

\newcommand{\VL}{\left( \begin{array}{c}}
\newcommand{\VR}{\end{array} \right)}
\newcommand{\ML}{\left( \begin{array}{cc}}
\newcommand{\MLd}{\left( \begin{array}{ccc}}
\newcommand{\MLv}{\left( \begin{array}{cccc}}
\newcommand{\MR}{\end{array} \right)}

\definecolor{Lightblue}{cmyk}{0.9,0.1,0.1,0.3}
\definecolor{dgelborange}{cmyk}{0.,0.3,0.5, 0.}
\definecolor{Orange}{cmyk}{0.,0.5,0.5, 0.}
\definecolor{Lila}{rgb}{0.5,0.,1}

\newcommand{\complex}{{{\rm I} \kern -.59em {\rm C}}}

\newcommand{\nn}{\nonumber}

\begin{document}

\thispagestyle{empty}

\title{
\vspace*{-3.5cm} 
\hfill \normalsize Dedicated to the memory of Martinus Veltman,\\
\hfill a great physicist and among the main founders\\
\hfill of the standard model of particle physics\\ \vspace{0.5cm}
\LARGE From Veltman's conditions to Finite Unification}
\date{}
\author{ \hspace*{7mm} M. Mondrag\'on$^1$\thanks{email: myriam@fisica.unam.mx} ,
G. Patellis$^2$\thanks{email: patellis@central.ntua.gr}
and G. Zoupanos$^{2,3,4}$\thanks{email: George.Zoupanos@cern.ch}\\
{\small
$^1$Instituto de F\'{\i}sica, Universidad Nacional Aut\'onoma de M\'exico, A.P. 20-364, CDMX 01000 M\'exico\\ %
$^2$ Physics Department,   Nat. Technical University, 157 80 Zografou, Athens, Greece\\
$^3$ Max-Planck Institut f\"ur Physik, F\"ohringer Ring 6, D-80805
  M\"unchen, Germany \\
$^4$ Theoretical Physics Department, CERN, Geneva, Switzerland \\
}
}
\maketitle

\abstract{
 First we review Veltman's suggestion to attack the naturalness problem in the Standard model by requiring absence of quadratic divergences and the resulting mass formula. Then we emphasise the influence of Veltman's suggestion in strengthening the belief that supersymmetry is the natural playground for solving the problem of quadratic divergences. Going further, we recall few sporadic suggestions concerning the cancellation of the logarithmic divergences too, which in the framework of supersymmetry has led to the construction of all-loop Finite Theories with the use of the idea of reduction of couplings. Eventually, we concentrate on a specific Finite Unified Theory and its successful predictions for the top and Higgs mass, among others, and the prospects of its final justification in future collider searches.
}


\section{Introduction}

Tini Veltman was a great scientific personality who contributed continuously
for decades in the most significant manner in establishing what everybody
accepts today as the Standard Model (SM) of Elementary Particle Physics.\footnote{A more personal, but also more detailed presentation of Veltman's
contributions together with some biografical notes can be found at the Corfu Institute (EISA) homepage:
http://eisa.institute .}

His numerous important contributions in the field and in particular the
breakthrough of Tini Veltman and Gerard ‘t Hooft on the renormalizability of the Standard Model (SM), one of the great moments in twentieth century physics, was awarded a Nobel
Prize in Physics in 1999 “for elucidating the quantum structure of electroweak
interactions in physics”.

Here, we would like to present how one of Tini Veltman's ideas influenced the
development of a particular direction of research, which eventually led to early successful 
predictions of the top quark and Higgs masses.

A concept that inspired Veltman in the direction that we would like to discuss
and which is in the center of theoretical discussions after Veltman's work is
the \textit{naturalness}  of a theory \cite{Wilson:1970ag,Susskind:1978ms,tHooft:1980xss}. According to this idea, a theory is
considered natural if at ordinary energies it is not too sensitive to the
fundamental constants of nature. More specifically, a theory is considered
unnatural if the radiative corrections to a physical observable have an
intrinsic magnitude much greater than the observed value, so that a conspiracy
among different orders in perturbation theory or a ``fine tuning'' is required.
The naturalness criterion is particularly serious in the case of the SM since it
belongs to the general category of renormalizable field theories with scalar
masses which are known to suffer from quadratic divergences. Then quadratic
divergences are indicative of the fact that the natural order of magnitude of
the Higgs mass in the SM is $\mathcal{O}(f_L \Lambda)$, where $f_L$ is a loop factor and $\Lambda$ is
the scale of new physics beyond the SM. Clearly then, absence of quadratic
divergences is a necessary condition for the naturalness of the SM, which has to
be modified in such a way so they are removed, and that the mass scale of
the modification
should be in the TeV scale. This requirement is not sufficient, since such a theory might still
suffer from the gauge hierarchy problem, i.e. could not provide the reason
that there exist scales with huge differences in magnitude in nature, as
for instance among the electroweak and the Planck scale.

With considerations along the above lines, Veltman was led to impose the
condition of absence of the quadratic divergences in the SM in his famous paper
published in Acta Physics Polonica \cite{Veltman:1980mj}. It is a very important work since it was
shown that this condition, known as the Veltman condition, is not just
technical, but was leading to a relation among the masses of the SM, i.e. it has
physical consequences, although after the discovery of the top and Higgs particles it
appeared not to hold. Equally important is the fact that this work paved the way for
supersymmetry (SUSY) \cite{Wess:1973kz}  to be considered widely, and not only among the experts, as a theory
with physical significance and consequences. Specifically, renormalizable
supersymmetric theories are free of quadratic divergences to all orders in
perturbation theory due to non-renormalization theorems. Of particular interest
is the fact that such a property holds also in theories with softly broken
supersymmetry (SSB) \cite{Iliopoulos:1974zv,Wess:1974jb,Ferrara:1975ye,Girardello:1981wz}, such as the celebrated
Minimal Supersymmetric Standard Model (MSSM) (for details see \cite{Martin:1997ns}), which has good chances to
describe Physics beyond the SM. Finally, it should be stressed that for a very
general class of theories with spontaneously broken supersymmetry, a mass formula
was derived \cite{Ferrara:1979wa}, which is very similar to the one resulting from Veltman's condition
for cancellation of quadratic divergences.

The next question in this exciting avenue of development of ideas, starting from
Veltman's fundamental work in \cite{Veltman:1980mj}, concerns the uniqueness of supersymmetry as a
solution to the problem of cancellation of quadratic divergences in
renormalizable field theories involving scalars. This was posed by two groups in
 \cite{Inami:1982xb} and \cite{Deshpande:1983ka,Deshpande:1984ke} and was answered positively. Indeed, supersymmetry is the
unique way to cancel the quadratic divergences in renormalizable field theories
with scalars that can be examined perturbatively. Still, there is another very
interesting way to avoid the problem by considering that the scalars are not
fundamental but composite, i.e. a bound state of two fermions and was also
mentioned in Veltman's paper in ref \cite{Veltman:1980mj}.

It should also be noted that Decker and Pestieau did, independently of Veltman, a
similar analysis but they went a step further requiring that the lepton
self-masses be \textit{finite} \cite{Lucchesi:1987he}, i.e. cancellation of the logarithmic
divergences, too. As a result, new mass relations were found.
 Inspired by all the above ideas, we were searching for the construction of
realistic Finite Theories with predictive power concerning some of the SM free
parameters, the proliferation of which was always considered as another big
obstacle of this theory. Quite naturally we were led to the framework of SSB
supersymmetric theories where cancellation of quadratic divergences holds to all
orders in perturbation theory and moreover to require absence of logarithmic
divergences. It is remarkable that all-orders finite supersymmetric gauge
theories can be constructed using the \textit{reduction of couplings} scheme
\cite{Lucchesi:1987he} and we consider ourselves lucky that we managed to construct the first
realistic Finite Unified Theory \cite{Kapetanakis:1992vx,Mondragon:1993tw}. Moreover, this model was predicting
correctly the top quark mass one and a half year before its discovery; a
prediction which survived for twelve years. Another version of the model \cite{Kobayashi:1997qx}  was predicting -in
addition to the top quark mass- the Higgs boson mass, four and half years
before the experimental discovery \cite{Heinemeyer:2007tz}.
 
 In the present paper we present in \refse{veltman} briefly some details on Veltman's
condition on the cancellation of quadratic divergences in SM with comments of
other authors and similarly the works on the uniqueness of  supersymmetry as a
solution to the cancellation of quadratic divergences problem. Then we continue
in \refse{roc} with the presentation of the scheme of reduction of couplings and \refse{finiteness} with the necessary conditions for finiteness. In \refse{futb} we review the above-mentioned all-loop finite $N=1$ supersymmetric $SU(5)$ model and give its latest phenomenological analysis. \refse{conclusions} is dedicated to a few closing remarks.

\section{Cancellation of quadratic divergences and Supersymmetry}\label{veltman}

\subsection{Comments on Veltman's relation}

Let us present few more details on Veltman's relation resulting from the
requirement of cancellation of quadratic divergences in the SM at one loop.
Veltman suggested that within the dimensional regularization \cite{tHooft:1972tcz}, which
does not catch the quadratic divergences, a suitable criterion of
identifying such divergences is the occurence of poles in the complex
dimensional plane of $n$ less than four.
Therefore quadratic divergences at the
one loop level would correspond to poles for $n=2$.
Then, in the SM within the dimensional regularization scheme poles for n=2 occur
in the vector boson and Higgs self energy and in the tadpole diagrams. However 
Veltman, inspired by the way dimensional regularisation has to be modified in
order for the scheme to be suitable also for supersymmetric theories \cite{Siegel:1979wq,Capper:1979ns}, 
chose the dimension of the Dirac matrices to be four, independent of the
space-time dimension. In other words, Veltman concluded that although
conventional dimensional regularisation would suggest $n = 2$ as the dimension of the Dirac algebra,
the appropriate choice is $n=4$. This preserves the number of gauge degrees of freedom and hence
respects supersymmetry, and corresponds to the use of regularisation by
dimensional reduction \cite{Siegel:1979wq,Capper:1979ns}. In any case, in ref.~\cite{Jack:1994bn},
the equivalence
of dimensional reduction and dimensional regularisation was shown.
With the above reasoning Veltman derived the following mass relation:
\begin{equation}
   m_e^2 +m_\mu^2 + m_\tau^2 + 3(m_u^2 +m_d^2 +m_c^2 +m_s^2 +m_t^2 +m_b^2)
 = \frac{3}{2} m_W^2 + \frac{3}{4}m_Z^2 + \frac{3}{4} m_H^2~, 
 \label{eq:Veltmansmass}
\end{equation}
known as Veltman's mass relation. It is very interesting that the same formula
was derived in ref.~\cite{Osland:1992zh}, based on the point-splitting regularization \cite{Osland:1992ay}, which
makes no reference to dimensions of space-time other than four. For discussions
concerning the two-loop corrections we refer the reader to \cite{Osland:1992cq,Alsarhi:1991ji}. Clearly now, given
the measured values of the top and Higgs masses, the relation (\ref{eq:Veltmansmass}) does not
hold.
  From the above discussion it is worth keeping the point that Veltman, although
working within the SM had a vision that supersymmetric theories were the
appropriate framework for the cancellation of quadratic divergences, in which
the contributions of the bosons and the fermions have opposite sign. Equally
important is the observation that the requirement of the cancellation of
quadratic divergences was leading to very useful mass relations. Actually a
strong support towards this direction
was already provided by the work of Ferrara, Girardello and Palumbo \cite{Ferrara:1979wa}, who
derived in supersymmetric theories with spontaneously broken supersymmetry
a very similar quadratic mass formula:
\begin{equation}
   \sum_J (-1)^{2J}(2J+1) m_J^2 = 0. 
   \label{eq:FerraraGirPal}
\end{equation}
                      
Some further interesting comments on Veltman's relation were done by Kubo,
Sibold and Zimmermann \cite{Kubo:1988jc} using the reduction of couplings scheme, which will
be discused in detail in the next section. Here we would like only to remind
that based on this scheme the parameters of SM were related to $\alpha_s$ leading
to predictions for the top and Higgs masses \cite{Kubo:1985up,Kubo:1988zu} that do not hold. In ref.~\cite{Kubo:1988jc}
it was analyzed whether it is possible to require in addition the absence of
quadratical divergences using Veltman's relation. It has been shown first that
postulating absence of quadratical divergences is a gauge and renormalization
group invariant statement. Moreover the resulting constraint is compatible with
reduction, at least with what they called 'the trivial one', meaning that the top
mass was considered as another free parameter instead of been predicted by the
dimensional reduction as in \cite{Kubo:1985up,Kubo:1988zu}.

\subsection{Uniqueness of supersymmetry as solution of the quadratic divergences problem}

The absence of quadratic divergences in supersymmetric theories was known due
to the non-renormalization theorems, as is already mentioned in the Introduction.
A further question was if this solution was unique, that is, if there
are non-supersymmetric theories where also the quadratic divergences
are absent.

In order to address this problem, the inverse question was posed,
given the absence of quadratic divergences, what kind of solutions
does it imply.

The first development in this direction, following the spirit of
\cite{Veltman:1980mj}, was done in \cite{Inami:1982xb}.  In order to
show whether the absence of quadratic divergences implies
supersymmetry, they studied the case of one Majorana field and an
arbitrary number of scalar and pseudoscalar fields systematically.  
By carrying out the loop expansion to two-loop order, and requiring that
the quadratic divergences cancel order by order, they derived
relationships among the dimensionless couplings.
They found no solutions with only one spin 0 field, either scalar or
pseudoscalar. In case of either a pair of scalars or pseudoscalar
fields, the only solutions are the trivial ones.  The combination of
one scalar and one pseudoscalar gives a non-trivial solution to the
cancellation of quadratic divergences, which corresponds 
to the massive Wess-Zumino model with soft supersymmetry breaking terms.

They concluded that the necessary and sufficient conditions for the
absence of quadratic divergences to two-loop order in cases of two or less
bosonic fields lead uniquely to the softly broken supersymmetric
theories.  The statement can be extended to all-loops,
since in theories with soft breaking terms there will appear no
quadratic divergences.

Shortly later, in \cite{Deshpande:1983ka},  
the requirement of
absence of quadratic divergences in a quantum field theory with one
or more scalar bosons was studied.  The analysis was done at one-loop, but it was
further required that the constraints resulting from eliminating 
the quadratic divergences should 
be renormalization group invariant (RGI), in order to have a physical
meaning. More specifically, each scalar boson has associated a quadratic
divergence and demanding that these are eliminated leads to
parametric conditions, which then were required to be preserved under a
change of the renormalization scale. In general the resulting systems are
severely overconstrained. With a procedure very close to the reduction of
couplings (see next section) they determined the independent couplings
and the relations among them. In all cases considered in that paper, the
only solutions found were supersymmetric.

This work was extended and detailed in \cite{Deshpande:1984ke}, where
they considered general classes of theories with scalars, Abelian and 
non-Abelian, with quartic and Yukawa interactions. By requiring
renormalization group invariance at one-loop, besides the cancellation
of quadratic divergences, they found that the only solutions possible
are supersymmetric. As a notable exception they found that in a chiral supersymmetric $U(1)$
model the quadratic divergence associated with a radiatively induced
Fayet-Iliopoulos D-term, do not cancel. It seems in their analysis they missed that the
Veltman relation is RGI according to ref.~\cite{Kubo:1988jc}.

Thus, the absence of quadratic divergences induced by scalar
couplings, leads in general to a supersymmetric solution. Given that  SSB terms are by
construction free of quadratic divergences, the necessity to add them in
any supersymmetric model in the prospect to become realistic is very
welcome without any cost. 

\section{Theoretical Basis of Reduction of Couplings}\label{roc}

The idea of reduction of couplings was introduced in \cite{Zimmermann:1984sx} and evolved over the next~two decades. It aims to express the parameters of a theory -that are considered independent- in terms of one basic parameter, which is called primary~coupling. This is achieved by searching for Renormalization Group Invariant (RGI) relations among couplings and using them to reduce~the -seemingly- free parameters.
In this section we will outline the procedure, first  applied to parameters without mass~dimension, and then it will be extended to parameters~of dimension one or two, i.e. the parameters of the soft breaking~sector of an $N=1$ SUSY theory.

\subsection{Reduction of Dimensionless~Parameters}\label{roc_0}

Any RGI~relation among couplings $g_1,...,g_A$ of a renormalizable theory  can be written in the form $\Phi (g_1,\cdots,g_A) ~=~\mbox{const.}$, which has to satisfy~the  partial differential equation
\beq
\mu\,\frac{d \Phi}{d \mu} = {\vec \nabla}\Phi\cdot {\vec \beta} ~=~
\sum_{a=1}^{A}
\,\beta_{a}\,\frac{\partial \Phi}{\partial g_{a}}~=~0~,
\eeq
where $\beta_a$ is the $\beta$-function of $g_a$.
Solving this partial differential equation is equivalent to solving a set of ordinary differential equations, known as reduction~equations (REs) \cite{Zimmermann:1984sx,Oehme:1984yy,Oehme:1985jy},
\beq
\beta_{g} \,\frac{d g_{a}}{d g} =\beta_{a}~,~a=1,\cdots,A~,
\label{redeq}
\eeq
where $g$ and~$\beta_{g}$ are the  primary
coupling and its $\beta$-function, respectively, 
while the~counting on $a$ does not~include $g$.
Since the  $\Phi_a$'s can impose a maximum of  ($A-1$)~independent
RGI~``constraints''~in the~$A$-dimensional~space of parameters,
one~could
express them all in terms of
a single coupling $g$. 
However, the general solutions of \refeqs{redeq} contain as many integration constants as~the number of equations. Thus, we have just~traded an integration constant for~each renormalized coupling and such general solutions cannot be considered~``reduced ones''.~The  crucial requirement is to demand~power series~solutions to the REs which~preserve perturbative renormalizability,
\beq
g_{a} = \sum_{n}\rho_{a}^{(n)}\,g^{2n+1}~,
\label{powerser}
\eeq
This ansatz fixes the integration constant in each~of the REs and chooses a special solution. Remarkably, the  uniqueness of these power series~solutions can be decided already at one-loop~level
\cite{Zimmermann:1984sx,Oehme:1984yy,Oehme:1985jy}. As an illustration, we assume $\beta$-functions of~the form
\beq
\begin{split}
\beta_{a} &=\frac{1}{16 \pi^2}\left[ \sum_{b,c,d\neq
  g}\beta^{(1)\,bcd}_{a}g_b g_c g_d+
\sum_{b\neq g}\beta^{(1)\,b}_{a}g_b g^2\right]+\cdots~,\\
\beta_{g} &=\frac{1}{16 \pi^2}\beta^{(1)}_{g}g^3+ \cdots~.
\end{split}
\eeq
$\cdots$ stands~for higher-order terms, and $\beta^{(1)\,bcd}_{a}$'s
are symmetric~in $ b,c,d$.  We will assume that the  $\rho_{a}^{(n)}$'s with $n\leq r$ are uniquely determined. To obtain~$\rho_{a}^{(r+1)}$'s we insert the~power series (\ref{powerser}) into the~REs (\ref{redeq}) and collect~terms of ${\cal O}(g^{2r+3})$:
\beq
\sum_{d\neq g}M(r)_{a}^{d}\,\rho_{d}^{(r+1)} = \mbox{lower
  order~quantities}~,\non
\eeq
where the  right-hand~side is known by assumption~and
\begin{align}
M(r)_{a}^{d} &=3\sum_{b,c\neq
  g}\,\beta^{(1)\,bcd}_{a}\,\rho_{b}^{(1)}\,
\rho_{c}^{(1)}+\beta^{(1)\,d}_{a}
-(2r+1)\,\beta^{(1)}_{g}\,\delta_{a}^{d}~,\label{M}\\
0 &=\sum_{b,c,d\neq g}\,\beta^{(1)\,bcd}_{a}\,
\rho_{b}^{(1)}\,\rho_{c}^{(1)}\,\rho_{d}^{(1)} +\sum_{d\neq
  g}\beta^{(1)\,d}_{a}\,\rho_{d}^{(1)}
-\beta^{(1)}_{g}\,\rho_{a}^{(1)}~.
\end{align}
Therefore, the~$\rho_{a}^{(n)}$'s for~all $n > 1$ for~a
given set of $\rho_{a}^{(1)}$'s are uniquely~determined if $\det M(n)_{a}^{d} \neq 0$ for all $n \geq 0$. 

The couplings in SUSY theories have the same asymptotic behaviour. Thus, it is natural to  search for such a power series solution to the REs.
The prospect of coupling unification described in this section~is very attractive, as the  ``completely reduced'' theory contains~only one independent coupling, with primary examples the FUTs \cite{Kapetanakis:1992vx,Mondragon:1993tw,Kobayashi:1997qx}. However, since it is often unrealistic, one usually imposes fewer RGI~constraints, achieving  ``partial reduction'' \cite{Kubo:1985up,Kubo:1988zu}.\\

All the above hint (recall also \cite{Kubo:1988jc}) towards an underlying~connection among reduction of couplings and supersymmetry.  
As an~example, consider an $SU(N)$ gauge theory with $\phi^{i}({\bf N})$~and $\hat{\phi}_{i}(\overline{\bf N})$ complex scalars, $\psi^{i}({\bf N})$~and $\hat{\psi}_{i}(\overline{\bf N})$ left-handed~Weyl spinors
and $\lambda^a~(a=1,\dots,N^2-1)$ right-handed Weyl spinors in the adjoint representation of~$SU(N)$, i.e. a model with the field content of a supersymmetric theory, but not
with the corresponding couplings.
The Lagrangian then includes
\beq
{\cal L} \supset 
i \sqrt{2} \{~g_Y\overline{\psi}\lambda^a T^a \phi
-\hat{g}_Y\overline{\hat{\psi}}\lambda^a T^a \hat{\phi}
+\mbox{h.c.}~\}-V(\phi,\overline{\phi}),
\eeq
where
\beq
V(\phi,\overline{\phi}) =
\frac{1}{4}\lambda_1(\phi^i \phi^{*}_{i})^2+
\frac{1}{4}\lambda_2(\hat{\phi}_i \hat{\phi}^{*~i})^2
+\lambda_3(\phi^i \phi^{*}_{i})(\hat{\phi}_j \hat{\phi}^{*~j})+
\lambda_4(\phi^i \phi^{*}_{j})
(\hat{\phi}_i \hat{\phi}^{*~j})~.
\eeq
This is the most general~renormalizable form in 4D. Searching for a solution like those in \refeq{powerser} for the REs, one finds
among the many possible solutions in lowest order:
\beq
\begin{split}
g_{Y}&=\hat{g}_{Y}=g~,\\
\lambda_{1}&=\lambda_{2}=\frac{N-1}{N}g^2~,\\
\lambda_{3}&=\frac{1}{2N}g^2~,~
\lambda_{4}=-\frac{1}{2}g^2~,
\end{split}
\eeq
which corresponds~to a $N=1$ SUSY~gauge theory. While~the above do~not provide~an answer about the relation~of reduction~of couplings~and SUSY, they indeed point to further~study~in that direction.

\subsection{Reduction~of Couplings in \texorpdfstring{$N=1$}{Lg} SUSY Gauge~Theories - Partial Reduction}\label{roc_susy}

Let us consider a~chiral, $N=1$ supersymmetric
gauge theory with group G and gauge coupling $g$.~The~superpotential of the theory can be written:
\bea
W&=& \frac{1}{2}\,m_{ij} \,\phi_{i}\,\phi_{j}+
\frac{1}{6}\,C_{ijk} \,\phi_{i}\,\phi_{j}\,\phi_{k}~,
\label{supot0}
\eea
where $m_{ij}$ and~$C_{ijk}$ are gauge invariant tensors and~the chiral~superfield $\phi_{i}$ belongs to the irreducible~representation~$R_{i}$ of the gauge group. The renormalization constants~associated with the~superpotential, for~preserved SUSY, are:
\begin{align}
\phi_{i}^{0}&=\left(Z^{j}_{i}\right)^{(1/2)}\,\phi_{j}~,~\\
m_{ij}^{0}&=Z^{i'j'}_{ij}\,m_{i'j'}~,~\\
C_{ijk}^{0}&=Z^{i'j'k'}_{ijk}\,C_{i'j'k'}~.
\end{align}

\noindent By virtue~of the $N=1$ non-renormalization~theorem \cite{Wess:1973kz,Iliopoulos:1974zv,Ferrara:1974fv,Fujikawa:1974ay} there are~no mass and cubic interaction term infinities:
\begin{equation}
\begin{split}
Z_{ij}^{i'j'}\left(Z^{i''}_{i'}\right)^{(1/2)}\left(Z^{j''}_{j'}\right)^{(1/2)}
&=\delta_{(i}^{i''}
\,\delta_{j)}^{j''}~,\\
Z_{ijk}^{i'j'k'}\left(Z^{i''}_{i'}\right)^{(1/2)}\left(Z^{j''}_{j'}\right)^{(1/2)}
\left(Z^{k''}_{k'}\right)^{(1/2)}&=\delta_{(i}^{i''}
\,\delta_{j}^{j''}\delta_{k)}^{k''}~.
\end{split}
\end{equation}
Therefore, the only surviving infinities are
the wave~function renormalization constants $Z^{j}_{i}$, so just on infinity~per field. The one-loop $\beta$-function of $g$ is given~by
\cite{Parkes:1984dh,West:1984dg,Jones:1985ay,Jones:1984cx,Parkes:1985hh}
\beq
\beta^{(1)}_{g}=\frac{d g}{d t} =
\frac{g^3}{16\pi^2}\left[\,\sum_{i}\,T(R_{i})-3\,C_{2}(G)\right]~,
\label{betag}
\eeq
where~$C_{2}(G)$ is the quadratic~Casimir operator~of the adjoint~representation of the gauge group~$G$ and $\textrm{Tr}[T^aT^b]=T(R)\delta^{ab}$, where~$T^a$ are the group generators~in the appropriate~representation.
Due to the non-renormalization theorem \cite{Wess:1973kz,Iliopoulos:1974zv,Fujikawa:1974ay} 
the~$\beta$-functions of $C_{ijk}$
are related~to the
anomalous~dimension matrices $\gamma_{ij}$ of the matter~fields as:
\beq
\beta_{ijk} =
 \frac{d C_{ijk}}{d t}~=~C_{ijl}\,\gamma^{l}_{k}+
 C_{ikl}\,\gamma^{l}_{j}+
 C_{jkl}\,\gamma^{l}_{i}~.
\label{betay}
\eeq
The one-loop $\gamma^i_j$ is~given by \cite{Parkes:1984dh}:
\beq
\gamma^{(1)}{}_{j}^{i}=\frac{1}{32\pi^2}\,[\,
C^{ikl}\,C_{jkl}-2\,g^2\,C_{2}(R_{i})\delta^i_j\,],
\label{gamay}
\eeq
where~$C^{ijk}=C_{ijk}^{*}$. 

\noindent We~take $C_{ijk}$ to be real so that~$C_{ijk}^2$~are always~positive. The squares of the couplings~are~convenient to work with, and the $C_{ijk}$ can be covered~by a~single~index $i~(i=1,\cdots,n)$:
\beq
\alpha = \frac{g^2}{4\pi}~,~
\alpha_{i} ~=~ \frac{g_i^2}{4\pi}~.
\label{alfas}
\eeq
Then the~evolution~of $\alpha$'s in perturbation theory will take~the~form
\beq
\begin{split}
\frac{d\alpha}{d t}&=\beta~=~ -\beta^{(1)}\alpha^2+\cdots~,\\
\frac{d\alpha_{i}}{d t}&=\beta_{i}~=~ -\beta^{(1)}_{i}\,\alpha_{i}\,
\alpha+\sum_{j,k}\,\beta^{(1)}_{i,jk}\,\alpha_{j}\,
\alpha_{k}+\cdots~,
\label{eveq}
\end{split}
\eeq
Here, $\cdots$~denotes~higher-order contributions~and
$ \beta^{(1)}_{i,jk}=\beta^{(1)}_{i,kj} $.
For the~evolution~equations~(\ref{eveq}), following ref \cite{Kubo:1994bj} we investigate the
asymptotic~properties.~First, we~define
\cite{Zimmermann:1984sx,Oehme:1985jy,Oehme:1984iz,Cheng:1973nv,Chang:1974bv}
\beq
\tilde{\alpha}_{i} \equiv \frac{\alpha_{i}}{\alpha}~,~i=1,\cdots,n~,
\label{alfat}
\eeq
and~derive from~\refeq{eveq}
\beq
\begin{split}
\alpha \frac{d \tilde{\alpha}_{i}}{d\alpha} &=
-\tilde{\alpha}_{i}+\frac{\beta_{i}}{\beta}= \left(-1+\frac{\beta^{(1)}_{i}}{\beta^{(1)}}\,\right) \tilde{\alpha}_{i}\\
&
-\sum_{j,k}\,\frac{\beta^{(1)}_{i,jk}}{\beta^{(1)}}
\,\tilde{\alpha}_{j}\, \tilde{\alpha}_{k}+\sum_{r=2}\,
\left(\frac{\alpha}{\pi}\right)^{r-1}\,\tilde{\beta}^{(r)}_{i}(\tilde{\alpha})~,
\label{RE}
\end{split}
\eeq
where $\tilde{\beta}^{(r)}_{i}(\tilde{\alpha})~(r=2,\cdots)$
are~power series of $\tilde{\alpha}$'s and~can be~computed
from~the~$r^{th}$-loop $\beta$-functions.
We then~search~for~fixed points $\rho_{i}$ of \refeq{alfat} at $ \alpha = 0$. We have~to solve the~equation
\beq
\left(-1+\frac{\beta ^{(1)}_{i}}{\beta ^{(1)}}\right) \rho_{i}
-\sum_{j,k}\frac{\beta ^{(1)}_{i,jk}}{\beta ^{(1)}}
\,\rho_{j}\, \rho_{k}=0~,
\label{fixpt}
\eeq
assuming fixed~points of the~form
\beq
\rho_{i}=0~\mbox{for}~ i=1,\cdots,n'~;~
\rho_{i} ~>0 ~\mbox{for}~i=n'+1,\cdots,n~.
\eeq
Next, we~treat~$\tilde{\alpha}_{i}$ with $i \leq n'$
as small perturbations  to the undisturbed~system (defined by setting~$\tilde{\alpha}_{i}$  with $i \leq n'$ equal~to zero). It~is possible~to verify the
existence of~the unique~power~series solution of the reduction equations (\ref{RE}) to~all orders already at one-loop~level~\cite{Zimmermann:1984sx,Oehme:1984yy,Oehme:1985jy,Oehme:1984iz}:
\beq
\tilde{\alpha}_{i}=\rho_{i}+\sum_{r=2}\rho^{(r)}_{i}\,
\alpha^{r-1}~,~i=n'+1,\cdots,n~.
\label{usol}
\eeq
 These~are~RGI~relations among~parameters, and preserve formally perturbative~renormalizability.
So, in~the undisturbed system~there is only one~independent
parameter, the primary coupling $\alpha$.

The~non-vanishing $\tilde{\alpha}_{i}$~with $i \leq n'$ cause small~perturbations that enter in a~way that the reduced couplings~($\tilde{\alpha}_{i}$  with $i > n'$) become functions~both of~$\alpha$ and $\tilde{\alpha}_{i}$  with $i \leq n'$. Investigating such systems with partial reduction is very~convenient to~work with the following PDEs:
\beq
\begin{split}
\left\{ \tilde{\beta}\,\frac{\partial}{\partial\alpha}
+\sum_{a=1}^{n'}\,
\tilde{\beta_{a}}\,\frac{\partial}{\partial\tilde{\alpha}_{a}}\right\}~
\tilde{\alpha}_{i}(\alpha,\tilde{\alpha})
&=\tilde{\beta}_{i}(\alpha,\tilde{\alpha})~,\\
\tilde{\beta}_{i(a)}~=~\frac{\beta_{i(a)}}{\alpha^2}
-\frac{\beta}{\alpha^{2}}~\tilde{\alpha}_{i(a)}
&,\qquad
\tilde{\beta}~\equiv~\frac{\beta}{\alpha}~.
\end{split}
\eeq
These~equations~are~equivalent~to the REs (\ref{RE}), where, in order~to~avoid~any confusion, we let~$a,b$ run from $1$ to $n'$ and~$i,j$ from $n'+1$ to $n$. Then,~we search for solutions~of the~form
\beq
\tilde{\alpha}_{i}=\rho_{i}+
\sum_{r=2}\,\left(\frac{\alpha}{\pi}\right)^{r-1}\,f^{(r)}_{i}
(\tilde{\alpha}_{a})~,~i=n'+1,\cdots,n~,
\label{algeq}
\eeq
where $f^{(r)}_{i}(\tilde{\alpha}_{a})$ are power~series of $\tilde{\alpha}_{a}$.~The requirement~that~in the~limit of~vanishing~perturbations~we~obtain the~undisturbed~solutions (\ref{usol})~\cite{Kubo:1988zu,Zimmermann:1993ei} suggests this~type of~solutions. Once more, one can obtain  the conditions~for~uniqueness of $ f^{(r)}_{i}$ in~terms of~the lowest~order coefficients.

\subsection{Reduction of Dimension-1 and -2 Parameters}\label{roc_dim_1-2}

The extension of~the reduction of couplings method to massive parameters is~not~straightforward, since the technique~was originally aimed~at massless~theories~on~the basis~of
the Callan-Symanzik equation~\cite{Zimmermann:1984sx,Oehme:1984yy}. Many requirements~have to be met, such  as the normalization conditions~imposed on irreducible Green's functions \cite{Piguet:1989pc},~etc. ~significant progress has been made~towards this~goal,~starting from~\cite{Kubo:1996js}, where, as an~assumption, a~mass-independent renormalization scheme~renders~all RG~functions only trivially dependent on dimensional parameters.~Mass parameters can~then~be introduced~similarly to~couplings. 

This~was~justified~later~\cite{Breitenlohner:2001pp,Zimmermann:2001pq},~where it~was demonstrated that, apart from dimensionless~parameters,~pole~masses and gauge couplings, the model can~also include~couplings carrying a dimension and masses.~To~simplify the~analysis, we follow \citere{Kubo:1996js} and use~a~mass-independent renormalization scheme~as~well.

Consider a~renormalizable theory~that contains $(N + 1)$
dimension-0 couplings,~$\left(\hat g_0,\hat g_1, ...,\hat g_N\right)$, $L$~parameters~with~mass dimension-1, $\left(\hat h_1,...,\hat h_L\right)$,~and  $M$ parameters~with~mass dimension-2,~$\left(\hat m_1^2,...,\hat m_M^2\right)$.~The~renormalized~irreducible vertex~function $\Gamma$ satisfies the~RGE
\beq
\label{RGE_OR_1}
\mathcal{D}\Gamma\left[\Phi's;\hat g_0,\hat g_1, ...,\hat g_N;\hat h_1,...,\hat h_L;\hat m_1^2,...,\hat m_M^2;\mu\right]=0~,
\eeq
with
\beq
\label{RGE_OR_2}
\mathcal{D}=\mu\frac{\partial}{\partial \mu}+
\sum_{i=0}^N \beta_i\frac{\partial}{\partial \hat g_i}+
\sum_{a=1}^L \gamma_a^h\frac{\partial}{\partial \hat h_a}+
\sum_{\alpha=1}^M \gamma_\alpha^{m^2}\frac{\partial}{\partial \hat m_\alpha ^2}+
\sum_J \Phi_I\gamma^{\phi I}_{\,\,\,\, J}\,\frac{\delta}{\delta\Phi_J}~,
\eeq
where~$\beta_i$ are the $\beta$-functions of the dimensionless~couplings~$g_i$ and $\Phi_I$  are
the~matter~fields.~The mass,~trilinear~coupling and wave~function anomalous~dimensions,~respectively, are denoted by $\ga_\alpha^{m^2}$, $\ga_a^h$ and $\ga^{\phi I}_{\,\,\,\, J}$~ and $\mu$~denotes the energy scale.~For a~mass-independent~renormalization~scheme, the $\gamma$'s are given by
\beq
\label{gammas}
\begin{split}
\gamma^h_a&=\sum_{b=1}^L\gamma_a^{h,b}(g_0,g_1,...,g_N)\hat h_b,\\
\gamma_\alpha^{m^2}&=\sum_{\beta=1}^M \gamma_\alpha^{m^2,\beta}(g_0,g_1,...,g_N)\hat m_\beta^2+
\sum_{a,b=1}^L \gamma_\alpha^{m^2,ab}(g_0,g_1,...,g_N)\hat h_a\hat h_b~.
\end{split}
\eeq
The $\gamma_a^{h,b}$,~$\gamma_\alpha^{m^2,\beta}$ and $\gamma_\alpha^{m^2,ab}$~are power~series of the~(dimensionless)~$g$'s.

\vspace{0.35cm}

\noindent We search~for~a~reduced~theory~where
\[
g\equiv g_0,\qquad h_a\equiv \hat h_a\quad \textrm{for $1\leq a\leq P$},\qquad
m^2_\alpha\equiv\hat m^2_\alpha\quad \textrm{for $1\leq \alpha\leq Q$}
\]
are~independent parameters.~The reduction~of the rest of the parameters, namely
\beq
\label{reduction}
\begin{split}
\hat g_i &= \hat g_i(g), \qquad (i=1,...,N),\\
\hat h_a &= \sum_{b=1}^P f_a^b(g)h_b, \qquad (a=P+1,...,L),\\
\hat m^2_\alpha &= \sum_{\beta=1}^Q e^\beta_\alpha(g)m^2_\beta + \sum_{a,b=1}^P k^{ab}_\alpha(g)h_ah_b,
\qquad (\alpha=Q+1,...,M)
\end{split}
\eeq
is~consistent~with the RGEs~(\ref{RGE_OR_1}) and (\ref{RGE_OR_2}). The~following relations~should~be~satisfied
\beq
\label{relation}
\begin{split}
\beta_g\,\frac{\partial\hat g_i}{\partial g} &= \beta_i,\qquad (i=1,...,N),\\
\beta_g\,\frac{\partial \hat h_a}{\partial g}+\sum_{b=1}^P \gamma^h_b\,\frac{\partial\hat h_a}{\partial h_b} &= \gamma^h_a,\qquad (a=P+1,...,L),\\
\beta_g\,\frac{\partial\hat m^2_\alpha}{\partial g} +\sum_{a=1}^P \gamma_a^h\,\frac{\partial\hat m^2_\alpha}{\partial h_a} +  \sum_{\beta=1}^Q \gamma_\beta ^{m^2}\,\frac{\partial\hat m_\alpha^2}{\partial m_\beta^2} &= \gamma_\alpha^{m^2}, \qquad (\alpha=Q+1,...,M).
\end{split}
\eeq
Using~Eqs.~(\ref{gammas}) and~(\ref{reduction}),~they reduce to
\beq
\label{relation_2}
\begin{split}
&\beta_g\,\frac{df^b_a}{dg}+ \sum_{c=1}^P f^c_a\left[\gamma^{h,b}_c + \sum_{d=P+1}^L \gamma^{h,d}_c f^b_d\right] -\gamma^{h,b}_a - \sum_{d=P+1}^L \gamma^{h,d}_a f^b_d=0,\\
&\hspace{8.6cm} (a=P+1,...,L;\, b=1,...,P),\\
&\beta_g\,\frac{de^\beta_\alpha}{dg} + \sum_{\gamma=1}^Q e^\gamma_\alpha\left[\gamma_\gamma^{m^2,\beta} +
\sum _{\delta=Q+1}^M\gamma_\gamma^{m^2,\delta} e^\beta_\delta\right]-\gamma_\alpha^{m^2,\beta} -
\sum_{\delta=Q+1}^M \gamma_\alpha^{m^2,d}e^\beta_\delta =0,\\
&\hspace{8.3cm} (\alpha=Q+1,...,M ;\, \beta=1,...,Q),\\
&\beta_g\,\frac{dk_\alpha^{ab}}{dg}
+ 2\sum_{c=1}^P \left(\gamma_c^{h,a} + \sum_{d=P+1}^L \gamma_c^{h,d} f_d^a\right)k_\alpha^{cb}
+\sum_{\beta=1}^Q e^\beta_\alpha\left[\gamma_\beta^{m^2,ab} + \sum_{c,d=P+1}^L \gamma_\beta^{m^2,cd}f^a_cf^b_d \right.\\
&\left. +2\sum_{c=P+1}^L \gamma_\beta^{m^2,cb}f^a_c + \sum_{\delta=Q+1}^M \gamma_\beta^{m^2,d} k_\delta^{ab}\right]- \left[\gamma_\alpha^{m^2,ab}+\sum_{c,d=P+1}^L \gamma_\alpha^{m^2,cd}f^a_c f^b_d\right.\\
&\left. +2 \sum_{c=P+1}^L \gamma_\alpha^{m^2,cb}f^a_c + \sum_{\delta=Q+1}^M \gamma_\alpha^{m^2,\delta}k_\delta^{ab}\right]=0,\\
&\hspace{8cm} (\alpha=Q+1,...,M;\, a,b=1,...,P)~.
\end{split}
\eeq
The above relations~ensure~that the~irreducible vertex function of~the~reduced~theory
\beq
\label{Green}
\begin{split}
\Gamma_R&\left[\Phi\textrm{'s};g;h_1,...,h_P; m_1^2,...,m_Q^2;\mu\right]\equiv\\
&\Gamma \left[  \Phi\textrm{'s};g,\hat g_1(g)...,\hat g_N(g);
h_1,...,h_P,\hat h_{P+1}(g,h),...,\hat h_L(g,h);\right.\\
& \left.  \qquad\qquad\qquad m_1^2,...,m^2_Q,\hat m^2_{Q+1}(g,h,m^2),...,\hat m^2_M(g,h,m^2);\mu\right]
\end{split}
\eeq
has the~same~renormalization~group~flow as~the~original~one.

Assuming a~perturbatively~renormalizable reduced~theory, the functions $\hat g_i$, $f^b_a$, $e^\beta_\alpha$~and $k_\alpha^{ab}$ are~expressed~as power~series~in~the primary coupling:
\beq
\label{pert}
\begin{split}
\hat g_i & = g\sum_{n=0}^\infty \rho_i^{(n)} g^n,\qquad
f_a^b  =  g \sum_{n=0}^\infty \eta_a^{b(n)} g^n,\\
e^\beta_\alpha & = \sum_{n=0}^\infty \xi^{\beta(n)}_\alpha g^n,\qquad
k_\alpha^{ab}=\sum_{n=0}^\infty \chi_\alpha^{ab(n)} g^n.
\end{split}
\eeq
These~expansion coefficients~are found by~inserting~the above power series into~Eqs. (\ref{relation}), (\ref{relation_2}) and~requiring~the~equations to~be satisfied~at each~order of~$g$. It is~not~trivial to have a unique power series solution;~it~depends both on the theory and  the choice of independent~couplings.

If~there are no~independent~dimension-1 parameters ($\hat h$), their~reduction~becomes
\[
\hat h_a = \sum_{b=1}^L f_a^b(g)M,
\]
where~$M$ is a~dimension-1 parameter (i.e. a gaugino~mass, corresponding~to~the independent gauge coupling). If there
are no~independent dimension-2~parameters ($\hat m^2$), their reduction takes~the~form
\[
\hat m^2_a=\sum_{b=1}^M e_a^b(g) M^2.
\]

\subsection{Reduction of Couplings of Soft Breaking Terms in \texorpdfstring{$N=1$}~ SUSY Theories}\label{roc_soft}

The~reduction of~dimensionless couplings was~extended \cite{Kubo:1996js,Jack:1995gm}~to the SSB~dimensionful parameters~of $N=1$~supersymmetric theories. It was also found~\cite{Kawamura:1997cw,Kobayashi:1997qx} that~soft scalar~masses~satisfy~a universal~sum rule.\\
Consider the superpotential (\ref{supot0})
\beq
W= \frac{1}{2}\,\mu^{ij} \,\Phi_{i}\,\Phi_{j}+
\frac{1}{6}\,C^{ijk} \,\Phi_{i}\,\Phi_{j}\,\Phi_{k}~,
\label{supot-prime}
\eeq
and~the SSB~Lagrangian 
\beq
-{\cal L}_{\rm SSB} =
\frac{1}{6} \,h^{ijk}\,\phi_i \phi_j \phi_k
+
\frac{1}{2} \,b^{ij}\,\phi_i \phi_j
+
\frac{1}{2} \,(m^2)^{j}_{i}\,\phi^{*\,i} \phi_j+
\frac{1}{2} \,M\,\lambda_i \lambda_i+\mbox{h.c.}
\label{supot_l}
\eeq
The $\phi_i$'s~are the scalar~parts~of chiral~superfields $\Phi_i$, $\lambda$~are gauginos~and $M$ the unified gaugino mass.

The one-loop gauge and Yukawa beta-functions are given  by  (\ref{betag}) and (\ref{betay}), respectively, and the one-loop anomalous dimensions by (\ref{gamay}). 
We make~the~assumption that the REs admit~power~series solutions:
\beq
C^{ijk} = g\,\sum_{n=0}\,\rho^{ijk}_{(n)} g^{2n}~.
\label{Yg-prime}
\eeq
Since~we want~to obtain higher-loop results instead~of knowledge~of explicit~$\beta$-functions, we~require relations among~$\beta$-functions. The spurion~technique
\cite{Fujikawa:1974ay,Delbourgo:1974jg,Salam:1974pp,Grisaru:1979wc,Girardello:1981wz} gives~all-loop relations~among SSB~$\beta$-functions
\cite{Yamada:1994id,Kazakov:1997nf,Jack:1997pa,Hisano:1997ua,Jack:1997eh,Avdeev:1997vx,Kazakov:1998uj,Karch:1998qa}:
\begin{align}
\beta_M &= 2{\cal O}\left(\frac{\beta_g}{g}\right)~,
\label{betaM}\\
\beta_h^{ijk}&=\gamma^i_l h^{ljk}+\gamma^j_l h^{ilk}
+\gamma^k_l h^{ijl}\non\\
&\,-2\left(\gamma_1\right)^i_l C^{ljk}
-2\left(\gamma_1\right)^j_l C^{ilk}-2\left(\gamma_1\right)^k_l C^{ijl}~,\label{betah}\\
(\beta_{m^2})^i_j &=\left[ \Delta
+ X \frac{\partial}{\partial g}\right]\gamma^i_j~,
\label{betam2}
\end{align}
where
\begin{align}
{\cal O} &=\left(Mg^2\frac{\partial}{\partial g^2}
-h^{lmn}\frac{\partial}{\partial C^{lmn}}\right)~,
\label{diffo}\\
\Delta &= 2{\cal O}{\cal O}^* +2|M|^2 g^2\frac{\partial}
{\partial g^2} +\tilde{C}_{lmn}
\frac{\partial}{\partial C_{lmn}} +
\tilde{C}^{lmn}\frac{\partial}{\partial C^{lmn}}~,\\
(\gamma_1)^i_j&={\cal O}\gamma^i_j,\\
\tilde{C}^{ijk}&=
(m^2)^i_l C^{ljk}+(m^2)^j_l C^{ilk}+(m^2)^k_l C^{ijl}~.
\label{tildeC}
\end{align}

\noindent Assuming (following~\cite{Jack:1997pa}) that~the relation~among couplings
\beq
h^{ijk} = -M (C^{ijk})'
\equiv -M \frac{d C^{ijk}(g)}{d \ln g}~,
\label{h2}
\eeq
is RGI to all orders and~the~use~of the all-loop~gauge $\beta$-function of \cite{Novikov:1983ee,Novikov:1985rd,Shifman:1996iy}
\beq
\beta_g^{\rm NSVZ} =
\frac{g^3}{16\pi^2}
\left[ \frac{\sum_l T(R_l)(1-\gamma_l /2)
-3 C_2(G)}{ 1-g^2C_2(G)/8\pi^2}\right]~,
\label{bnsvz}
\eeq
we are~led to an all-loop RGI sum rule~\cite{Kobayashi:1998jq} (assuming $(m^2)^i_j=m^2_j\delta^i_j$),
\begin{equation}
\begin{split}
m^2_i+m^2_j+m^2_k &=
|M|^2 \left\{~
\frac{1}{1-g^2 C_2(G)/(8\pi^2)}\frac{d \ln C^{ijk}}{d \ln g}
+\frac{1}{2}\frac{d^2 \ln C^{ijk}}{d (\ln g)^2}~\right\}\\
& \qquad\qquad +\sum_l
\frac{m^2_l T(R_l)}{C_2(G)-8\pi^2/g^2}
\frac{d \ln C^{ijk}}{d \ln g}~.
\label{sum2}
\end{split}
\end{equation}
It is worth noting~that~the all-loop result of \refeq{sum2} coincides with~the superstring result~for~the~finite case~in a certain~class of~orbifold~models \cite{Ibanez:1992hc,Brignole:1995fb,Kobayashi:1997qx}~if $\frac{d \ln C^{ijk}}{d \ln g}=1$~\cite{Mondragon:1993tw}.

As~mentioned~above, the~all-loop~results~on the~SSB~$\beta$-functions, Eqs.(\ref{betaM})-(\ref{tildeC}),
lead to~all-loop RGI~relations.~We assume:\\
(a)~the existence~of~an RGI~surface~on~which $C = C(g)$, or equivalently~that~the~expression
\beq
\label{Cbeta}
\frac{dC^{ijk}}{dg} = \frac{\beta^{ijk}_C}{\beta_g}
\eeq
holds~(i.e.~reduction~of couplings is~possible)\\
(b)~the~existence of a~RGI~surface on~which
\beq
\label{h2NEW}
h^{ijk} = - M \frac{dC(g)^{ijk}}{d\ln g}
\eeq
holds~to all~orders.\\
Then it~can~be proven~\cite{Jack:1999aj,Kobayashi:1998iaa,Kobayashi:2001me} that the relations~that~follow are all-loop~RGI (note that in
both assumptions~we~do not rely on specific~solutions of these~equations)
\begin{align}
M &= M_0~\frac{\beta_g}{g} ,  \label{M-M0} \\
h^{ijk}&=-M_0~\beta_C^{ijk},  \label{hbeta}  \\
b^{ij}&=-M_0~\beta_{\mu}^{ij},\label{bij}\\
(m^2)^i_j&= \frac{1}{2}~|M_0|^2~\mu\frac{d\gamma^i{}_j}{d\mu},
\label{scalmass}
\end{align}
where~$M_0$~is an arbitrary~reference~mass~scale~to be specified~shortly. Assuming
\beq
C_a\frac{\partial}{\partial C_a}
= C_a^*\frac{\partial}{\partial C_a^*} \label{dc/dc}
\eeq
for an~RGI surface $F(g,C^{ijk},C^{*ijk})$ we~are led~to
\begin{equation}
\label{F}
\frac{d}{dg} = \left(\frac{\partial}{\partial g} + 2\frac{\partial}{\partial C}\,\frac{dC}{dg}\right)
= \left(\frac{\partial}{\partial g} + 2 \frac{\beta_C}{\beta_g}
\frac{\partial}{\partial C} \right)\, ,
\end{equation}
where \refeq{Cbeta} was~used.~Let us now consider~the~partial differential operator~${\cal O}$ in
\refeq{diffo} which~(assuming~\refeq{h2}), becomes
\beq
{\cal O} = \frac{1}{2}M\frac{d}{d\ln g}\, 
\eeq
and~$\beta_M$, given in~\refeq{betaM}, becomes
\beq
\beta_M = M\frac{d}{d\ln g} \big( \frac{\beta_g}{g}\big) ~, \label{betaM2}
\eeq
which by integration~provides us \cite{Karch:1998qa,Jack:1999aj} with~the
generalized, i.e.~including Yukawa couplings,~all-loop RGI Hisano - Shifman~relation \cite{Hisano:1997ua}
\beq
 M = \frac{\beta_g}{g} M_0~.
\eeq
$M_0$ is~the integration~constant~and can~be associated~to the~unified gaugino mass $M$~(of an assumed covering GUT), or to~the~gravitino mass~$m_{3/2}$ in~a supergravity~framework. Therefore,~\refeq{M-M0} becomes the~all-loop RGI \refeq{M-M0}.~$\beta_M$, using~Eqs.(\ref{betaM2}) and (\ref{M-M0})~can be written as~follows:
\beq \beta_M =
M_0\frac{d}{dt} (\beta _g/g)~.
\label{eq:betaM}
\eeq
Similarly
\beq (\gamma_1)^i_j =
{\cal O} \gamma^i_j = \frac{1}{2}~M_0~\frac{d
  \gamma^i_j}{dt}~. \label{gammaO}
\eeq
Next,~from~Eq.(\ref{h2}) and Eq.(\ref{M-M0}) we get
\beq
 h^{ijk} = - M_0 ~\beta_C^{ijk}~,  \label{hm32}
\eeq
while~$\beta^{ijk}_h$,~using Eq.(\ref{gammaO}), becomes \cite{Jack:1999aj}
\beq
  \beta_h^{ijk} = - M_0~\frac{d}{dt} \beta_C^{ijk},
  \label{eq:beta_hijk}
\eeq
which~shows that~\refeq{hm32} is  RGI to all loops.~\refeq{bij} can similarly~be~shown to~be all-loop~RGI as well.

It should be noted concerning the $\beta$-functions of the SBB
parameters, as in Eqs.~(\ref{eq:betaM}) and (\ref{eq:beta_hijk}),  that the vanishing of the dimensionless
$\beta$-functions, even to all-orders, as will be discussed in the next
section, is transferred to the dimensionful SSB sector of the theory.

\section{Finiteness}\label{finiteness}

A natural development of the ideas started with Veltman's work on the
cancellation of quadratic divergences in renormalizable field theories
with scalars, which found an excellent realisation in supersymmetric
theories with soft supersymmetry breaking terms, as we have already
discussed, led to the search of constructing renormalizable field
theories free also of logarithmic divergences, i.e. completely \textit{Finite
Theories}.

The finiteness that will be discussed here is a consequence of the
reduction of couplings, presented in the previous section, and is based on
the fact that in supersymmetric theories is possible to find RGI
relations among couplings that keep finiteness in perturbation theory,
even to all orders.
Accepting~finiteness as a guiding~principle~in constructing realistic~theories of EPP, the~first~thing that comes to mind~is to look for an $N=4$~supersymmetric~unified gauge theory, since~any ultraviolet~(UV) divergences~are absent in these~theories. However~nobody~has managed so far to produce~realistic models in~the~framework of $N=4$~SUSY. In the best case one could~try to~do a drastic truncation of~the theory like the~orbifold~projection of~refs. \cite{Kachru:1998ys,Chatzistavrakidis:2010xi},~but this is already a~different~theory than the original~one. The next possibility~is~to consider an~$N=2$ supersymmetric gauge theory,~whose~$\beta$-function receives~corrections only at one loop.~Then~it is~not hard to select a~spectrum to make~the~theory~all-loop finite. However~a serious obstacle in these~theories~is their mirror~spectrum, which in the~absence~of a mechanism to make it~heavy, does not permit the construction~of realistic models.~Therefore, one is naturally~led to~consider $N=1$ supersymmetric gauge~theories,~which can be chiral~and in principle realistic.

It should~be noted~that in the approach followed~here (UV)~finiteness~means the vanishing of~all the $\beta$-functions, i.e.~the non-renormalization~of the coupling constants, in~contrast to a complete~(UV) finiteness where even~field~amplitude renormalization is~absent.
Before~the work of~several members of our group,~the studies on $N=1$~finite~theories were following two~directions: (i) construction~of finite~theories up to two loops examining various possibilities~to make them phenomenologically~viable, (ii) construction~of all-loop~finite models without particular~emphasis~on the phenomenological consequences. The success~of~the work of our group started in refs \cite{Kapetanakis:1992vx,Mondragon:1993tw} with the
construction of the first realistic all-loop~finite model, based on the theorem~presented~below, realising in this~way an old theoretical~dream of field~theorists.

\subsection*{Finiteness in N=1 Supersymmetric~Gauge Theories}

Let us,~once more,~consider a chiral,~anomaly free, $N=1$ globally~supersymmetric~gauge theory based on a group G with gauge~coupling~constant $g$.
The~superpotential~of the theory is given by (see \refeq{supot0})
\beq
W= \frac{1}{2}\,m_{ij} \,\phi_{i}\,\phi_{j}+
\frac{1}{6}\,C_{ijk} \,\phi_{i}\,\phi_{j}\,\phi_{k}~.
\label{supot}
\eeq
The~$N=1$ non-renormalization theorem, ensuring the absence of~mass and cubic-interaction-term~infinities, leads to wave-function~infinities only; one for each superfield.
As one~can~see from Eqs. (\ref{betag}) and~(\ref{gamay}),
all the~one-loop $\beta$-functions of the theory vanish if
$\beta_g^{(1)}$~and~$\gamma^{(1)}{}_{j}^{i}$ vanish, i.e.
\begin{align}
\sum _i T(R_{i})& = 3 C_2(G) \,,
\label{1st}     \\
 C^{ikl} C_{jkl} &= 2\delta ^i_j g^2  C_2(R_i)\,.
\label{2nd}
\end{align}
The conditions~for~finiteness for $N=1$ field theories with $SU(N)$ gauge~symmetry~are discussed in \cite{Rajpoot:1984zq},~and  the~analysis of the  anomaly-free and  no-charge~renormalization~requirements for these theories can be~found in \cite{Rajpoot:1985aq}.~A very interesting~result is that the  conditions (\ref{1st}) and (\ref{2nd})~are~necessary and  sufficient for finiteness at the~two-loop~level
\cite{Parkes:1984dh,West:1984dg,Jones:1985ay,Jones:1984cx,Parkes:1985hh}.

In case SUSY is broken by soft terms, the requirement of
finiteness in the one-loop~soft breaking terms imposes~further~constraints among them  \cite{Jones:1984cu}. In addition, the same set of conditions that are sufficient for one-loop finiteness of the soft breaking terms render the~soft sector of the theory two-loop~finite~\cite{Jack:1994kd}.

The one- and~two-loop~finiteness conditions of Eqs. (\ref{1st}) and (\ref{2nd})~restrict~considerably the  possible choices of the~irreducible~representations~(irreps)
$R_i$ for a~given~group $G$, as well as the  Yukawa couplings in the~superpotential~(\ref{supot}).~Note in particular~that the  finiteness~conditions cannot be~applied to the MSSM, since the~presence
of a $U(1)$ gauge~group~is incompatible with the~condition
(\ref{1st}), due~to $C_2[U(1)]=0$.~This naturally leads to the~expectation that finiteness should~be attained at the  grand~unified~level only, the MSSM being just the~corresponding,~low-energy,~effective theory.

Another~important~consequence of one- and two-loop finiteness is that~SUSY~(most probably) can only be broken~due to the  soft~breaking~terms. Indeed, due to the unacceptability of gauge~singlets, F-type spontaneous~symmetry~breaking \cite{ORaifeartaigh:1975nky}~terms~are incompatible with finiteness, as~well as~D-type \cite{Fayet:1974jb} spontaneous breaking which~requires~the  existence of a $U(1)$ gauge~group.

A natural~question to~ask is what happens at higher loop orders.~The answer is~contained in~a theorem
\cite{Lucchesi:1987he,Lucchesi:1987ef}~which~states the  necessary and~sufficient~conditions to achieve finiteness~at all orders.  Before~we~discuss the  theorem let us make some introductory~remarks.~The finiteness conditions impose relations~between~gauge and Yukawa couplings.~To require such relations which~render the  couplings~mutually dependent at a given~renormalization point is trivial.  What~is not trivial is to~guarantee that relations leading~to a reduction of the  couplings hold~at any renormalization point.~As we have seen
(see~\refeq{Cbeta}), the necessary~and also sufficient, condition for~this to happen is to~require that such relations are~solutions to the~REs
\beq \beta _g
\frac{d C_{ijk}}{dg} = \beta _{ijk}
\label{redeq2}
\eeq
and hold at all~orders.~Remarkably, the existence of~all-order~power series solutions to (\ref{redeq2})~can be decided at~one-loop level, as already~mentioned.

Let us now turn~to the~all-order finiteness theorem
\cite{Lucchesi:1987he,Lucchesi:1987ef},~which~states under which conditions an~$N=1$~supersymmetric gauge theory can become finite to~all~orders in perturbation theory, that is attain physical scale~invariance.  It is based on (a) the  structure of~the supercurrent in $N=1$ supersymmetric~gauge theory~\cite{Ferrara:1974pz,Piguet:1981mu,Piguet:1981mw}, and  on (b)~the~non-renormalization properties of~$N=1$ chiral anomalies~\cite{Lucchesi:1987he,Lucchesi:1987ef,Piguet:1986td,Piguet:1986pk,Ensign:1987wy}. Details~of the  proof can be found in~refs.~\cite{Lucchesi:1987he,Lucchesi:1987ef} and  further discussion~in~\citeres{Piguet:1986td,Piguet:1986pk,Ensign:1987wy,Lucchesi:1996ir,Piguet:1996mx}. Here,~following mostly~\citere{Piguet:1996mx} we~present a~comprehensible sketch of the~proof.

Consider~an~$N=1$ supersymmetric gauge theory, with simple Lie group~$G$.~The  content of this theory is~given at the  classical level~by~the matter supermultiplets $S_i$, which contain a scalar~field~$\phi_i$ and  a Weyl spinor $\psi_{ia}$, and~the~vector supermultiplet $V_a$, which contains a~gauge~vector field $A_{\mu}^a$ and a~gaugino
Weyl spinor~$\lambda^a_{\alpha}$.\\

\noindent Let~us first recall certain~facts about the  theory:

\noindent (1)~A~massless $N=1$ supersymmetric theory is invariant~under~a $U(1)$ chiral transformation $R$ under which the~various~fields transform as follows
\beq
\begin{split}
A'_{\mu}&=A_{\mu},~~\lambda '_{\alpha}=\exp({-i\theta})\lambda_{\alpha}\\
\phi '&= \exp({-i\frac{2}{3}\theta})\phi,~~\psi_{\alpha}'= \exp({-i\frac{1}
    {3}\theta})\psi_{\alpha},~\cdots
\end{split}
\eeq
The corresponding~axial~Noether current $J^{\mu}_R(x)$ is
\beq
J^{\mu}_R(x)=\bar{\lambda}\gamma^{\mu}\gamma^5\lambda + \cdots
\label{noethcurr}
\eeq
is conserved~classically,~while in the quantum case is violated by~the~axial anomaly
\beq
\partial_{\mu} J^{\mu}_R =
r\left(\epsilon^{\mu\nu\sigma\rho}F_{\mu\nu}F_{\sigma\rho}+\cdots\right).
\label{anomaly}
\eeq
From its~known~topological origin in ordinary~gauge theories
\cite{AlvarezGaume:1983cs,Bardeen:1984pm,Zumino:1983rz},~one would~expect the~axial vector current~$J^{\mu}_R$ to satisfy the Adler-Bardeen~theorem and~receive corrections only at the  one-loop~level. Indeed it has been shown that the same non-renormalization theorem~holds also in supersymmetric theories~\cite{Piguet:1986td,Piguet:1986pk,Ensign:1987wy}. ~Therefore
\beq
r=\hbar \beta_g^{(1)}.
\label{r}
\eeq

\noindent (2)~The~massless theory we consider is scale invariant at~the~classical level and, in general,~there is a scale anomaly due~to~radiative corrections.  The  scale anomaly~appears~in the  trace of the~energy momentum tensor $T_{\mu\nu}$,~which~is traceless classically.
It has the~form
\beq
T^{\mu}_{\mu} = \beta_g F^{\mu\nu}F_{\mu\nu} +\cdots
\label{Tmm}
\eeq

\noindent (3)~Massless,~$N=1$ supersymmetric gauge theories are~classically invariant~under the  supersymmetric extension of the~conformal~group -- the  superconformal group.  Examining the~superconformal~algebra, it can be seen that the  subset of~superconformal~transformations consisting of translations, SUSY~transformations, and  axial $R$ transformations is~closed under SUSY, i.e. these transformations form~a representation of SUSY.  It follows that the~conserved~currents~corresponding to these transformations make~up a supermultiplet represented by an axial~vector superfield called the supercurrent~$J$,
\beq
J \equiv \left\{ J'^{\mu}_R, ~Q^{\mu}_{\alpha}, ~T^{\mu}_{\nu} , ... \right\},
\label{J}
\eeq
where~$J'^{\mu}_R$ is~the  current associated to R invariance,~$Q^{\mu}_{\alpha}$~is the  one associated to SUSY invariance,~and $T^{\mu}_{\nu}$~the  one associated to translational~invariance~(energy-momentum tensor).\\
The anomalies of~the  R current $J'^{\mu}_R$, the  trace
anomalies of~the SUSY current, and  the  energy-momentum tensor, form~also a second supermultiplet, called the  supertrace~anomaly
\[
S =\left\{ Re~ S, ~Im~ S,~S_{\alpha}\right\}=
\left\{T^{\mu}_{\mu},~\partial _{\mu} J'^{\mu}_R,~\sigma^{\mu}_{\alpha
  \dot{\beta}} \bar{Q}^{\dot\beta}_{\mu}~+~\cdots \right\}
\]
where~$T^{\mu}_{\mu}$ is given in Eq.(\ref{Tmm}) and
\begin{align}
\partial _{\mu} J'^{\mu}_R &~=~\beta_g\epsilon^{\mu\nu\sigma\rho}
F_{\mu\nu}F_{\sigma \rho}+\cdots\\
\sigma^{\mu}_{\alpha \dot{\beta}} \bar{Q}^{\dot\beta}_{\mu}&~=~\beta_g
\lambda^{\beta}\sigma ^{\mu\nu}_{\alpha\beta}F_{\mu\nu}+\cdots
\end{align}

\noindent (4)~It is~very important to note that
the Noether~current~defined in (\ref{noethcurr}) is not the  same as the~current~associated to R invariance that appears in the~supercurrent~$J$ in (\ref{J}), but they coincide in the  tree~approximation. So starting from a unique classical Noether~current $J^{\mu}_{R(class)}$,  the  Noether
current~$J^{\mu}_R$~is defined as the  quantum extension of
$J^{\mu}_{R(class)}$ which~allows for the~validity of the  non-renormalization~theorem.~On the  other hand, $J'^{\mu}_R$, is~defined~to belong to the  supercurrent $J$,~together with~the  energy-momentum tensor.~The two requirements~cannot~be fulfilled by a single~current operator at the~same time.

Although the~Noether~current $J^{\mu}_R$ which obeys (\ref{anomaly})~and~the current $J'^{\mu}_R$ belonging to the  supercurrent~multiplet $J$ are not the  same, there is a relation \cite{Lucchesi:1987he,Lucchesi:1987ef} between quantities~associated~with them
\beq
r=\beta_g(1+x_g)+\beta_{ijk}x^{ijk}-\gamma_Ar^A
\label{rbeta}
\eeq
where $r$ was~given~in Eq.~(\ref{r}).  The  $r^A$ are the
non-renormalized coefficients~of the anomalies of the  Noether currents~associated to the chiral~invariances of the~superpotential, and --like $r$-- are strictly
one-loop~quantities.~The  $\gamma_A$'s are linear
combinations~of the~anomalous dimensions of the  matter fields,~and $x_g$, and  $x^{ijk}$ are radiative correction quantities. The~structure of Eq. (\ref{rbeta}) is independent of~the renormalization scheme.

One-loop~finiteness,~i.e. vanishing of the  $\beta$-functions at one-loop,~implies~that the  Yukawa couplings $\lambda_{ijk}$ must~be functions of~the gauge coupling $g$. To find a~similar condition to all orders it is necessary and sufficient~for~the Yukawa couplings to be a formal power series in~$g$,~which is solution of the~REs (\ref{redeq2}).

\noindent We can~now~state the theorem for all-order vanishing~$\beta$-functions \cite{Lucchesi:1987he}.
\bigskip

\noindent {\bf Theorem:}

\noindent Consider~an~$N=1$ supersymmetric Yang-Mills theory, with simple~gauge group. If the  following conditions~are satisfied
\begin{enumerate}
\item There~is no~gauge anomaly.
\item The~gauge~$\beta$-function vanishes at one-loop
  \beq
  \beta^{(1)}_g = 0 =\sum_i T(R_{i})-3\,C_{2}(G).
  \label{eq:1stfinitcond}
  \eeq
\item There~exist~solutions of the  form
  \beq
  C_{ijk}=\rho_{ijk}g,~\qquad \rho_{ijk}\in\complex
  \label{soltheo}
  \eeq
to the~conditions~of vanishing one-loop matter fields~anomalous dimensions
\beq
  \gamma^{(1)}{}_{j}^{i}~=~0
  =\frac{1}{32\pi^2}~[ ~
  C^{ikl}\,C_{jkl}-2~g^2~C_{2}(R)\delta_j^i ].
\eeq
\item These~solutions~are isolated and  non-degenerate when considered~as~solutions of vanishing one-loop~Yukawa $\beta$-functions:
   \beq
   \beta_{ijk}=0.
   \eeq
\end{enumerate}
Then, each of~the~solutions (\ref{soltheo}) can be~uniquely extended~to a~formal power series in $g$, and  the  associated super~Yang-Mills~models depend on the  single coupling constant~$g$ with a $\beta$-function~which vanishes at~all-orders.

\bigskip

It is~important~to note a few things: The~requirement of isolated and~non-degenerate
solutions~guarantees the~existence of a unique formal power series~solution to the reduction~equations.
The vanishing~of the  gauge $\beta$-function at one-loop, $\beta_g^{(1)}$,~is equivalent to the~vanishing of the  R current anomaly~(\ref{anomaly}).  The  vanishing of~the anomalous dimensions~at one-loop implies the  vanishing of the  Yukawa~couplings~$\beta$-functions at that order.  It also implies~the vanishing of the~chiral anomaly coefficients $r^A$. This~last property is a necessary~condition for having $\beta$-functions vanishing at all orders.\footnote{There~is an alternative~way to find finite~theories~\cite{Ermushev:1986cu,Kazakov:1987vg,Jones:1986vp,Leigh:1995ep}.}

\bigskip

\noindent {\bf Proof:}

\noindent Insert~$\beta_{ijk}$ as~given by the  REs into the relationship~(\ref{rbeta}).~Since these~chiral anomalies vanish, we get for $\beta_g$~an~homogeneous equation of the  form
\beq
0=\beta_g(1+O(\hbar)).
\label{prooftheo}
\eeq
The solution of~this equation in~the sense of a formal power series in~$\hbar$~is $\beta_g=0$, order by order.  Therefore, due to the~REs~(\ref{redeq2}), $\beta_{ijk}=0$ too.

Thus,~we see that~finiteness and  reduction of couplings are intimately~related.~Since an equation like eq.~(\ref{rbeta}) is lacking~in non-supersymmetric theories, one cannot extend the~validity of a~similar theorem in such theories.\\

A very~interesting development was done in~ref \cite{Kazakov:1997nf}. Based~on the all-loop relations among the $\beta$-functions of the~soft supersymmetry breaking terms~and those of the~rigid supersymmetric theory with the help of the differential~operators,~discussed in Sections \ref{roc_dim_1-2} and~\ref{roc_soft}, it was shown~that certain~RGI surfaces~can be chosen, so as to reach all-loop finiteness of~the full theory. More specifically it was shown~that on~certain RGI~surfaces the partial differential operators appearing~in~Eqs.~(\ref{betaM},\ref{betah}) acting on the~$\beta$- and~$\gamma$- functions of the rigid theory can be transformed to~total derivatives.~Then the all-loop finiteness of~the $\beta$- and $\gamma$-functions~of the rigid theory can~be transferred to the $\beta$-functions of the~soft~supersymmetry breaking terms. Therefore a~totally all-loop~finite $N=1$ SUSY gauge theory can be constructed, including~the soft supersymmetry breaking terms.

\section{Successful Finite Unification}\label{futb}

Below we briefly review the basic properties of a phenomenologically successful SUSY model with reduced couplings, which can be made finite to all-loops in perturbation theory. Its predictions for the top and bottom quark masses,  the SM Higgs boson mass, as well as the  supersymmetric and the other Higgs spectra are discussed in \ref{analysis}, while experimental constraints considered are listed in \ref{constraints}.  A few comments on Cold Dark Matter (CDM) are mentioned too.
 Other models with reduced~couplings that were analyzed in \cite{Heinemeyer:2020ftk} and~\cite{Heinemeyer:2020nzi} (see also \cite{Patellis:2021drd}~and \cite{Heinemeyer:2019vbc}) are the Reduced Minimal~$N=1$ $SU(5)$ \cite{Kubo:1994bj},  the two-loop Finite~$N=1$ $SU(3)^3$ \cite{Ma:2004mi,Heinemeyer:2010zza,Heinemeyer:2009zs}~and the Reduced Minimal Supersymmetric~Standard Model \cite{Mondragon:2013aea,Heinemeyer:2017gsv}.

\subsection{The Finite \texorpdfstring{$N=1$}{Lg} Supersymmetric \texorpdfstring{$SU(5)$}{Lg} Model}\label{model}

The model under review is a finite~to all-orders $SU(5)$ $N=1$ SUSY GUT (also referred to as {\bf FUTB}), where the finiteness conditions, resulting from the application of the reduction of couplings method and the requirement of vanishing one-loop $\beta$-functions, have been applied.

The particle content~of the model, resulting from applying condition (\ref{eq:1stfinitcond}), consists of three ($\overline{\bf 5} + \bf{10}$)~supermultiplets, where the quarks and leptons are accomodated,  while in the Higgs~sector there are four~supermultiplets~($\overline{\bf 5} + {\bf 5}$)~and one~${\bf 24}$.

The most general $SU(5)$ invariant,  cubic 
superpotential, where the R-parity that forbids fast proton decay has been imposed, and that is also consistent with the above particle content, is given by
\bea
W &=&H_{a}\,[~
f_{ab}\,\overline{H}_b {\bf 24}+ 
h_{ia}\,\overline{\bf 5}_i {\bf 24}
+\overline{g}_{ija}\,{\bf 10}_i \overline{\bf 5}_{j}]+
 p\,({\bf 24})^3 \nn\\
&+& \frac{1}{2}\,{\bf 10}_i\,[~
g_{ija}\,{\bf 10}_j H_a+
 \hat{g}_{iab}\,\overline{H}_a
\overline{H}_b+
g_{ijk}^{\prime}\,
\overline{\bf 5}_{j} \overline{\bf 5}_{k}~]~,
\eea
where $i,j,k=1,2,3$ and $a,b=1,\cdots,4$, and we sum over all
indices  in $W$ (notice that the $SU(5)$ indices are suppressed). 
The ${\bf 10}_{i}$'s
and $\overline{\bf 5}_{i}$'s are the usual 
three generations, and the four
$({\bf 5}+ \overline{\bf 5})$ Higgses are denoted by
 $H_a~,~\overline{H}_{a} $.  As further restrictions, to make the model viable, the anomalous dimensions have been assumed diagonal, and couplings between the fermions and the ${\bf 24}$ in the adjoint are not allowed. 
To achieve all-loop finiteness the  conditions 3 and 4 from the all-loop finiteness theorem have to be satisfied. These require the existence of isolated and non-degenerate solutions to the vanishing of the anomalous dimensions, and thus the vanishing of the Yukawa $\beta$-functions. One can check that this is indeed the case.  As explained in the previous section, these conditions guarantee a unique solution to the reduction equations.

The existence of these solutions implies an enhanced symmetry of the superpotential, which can be found e.g. in refs.~\cite{Kobayashi:1997qx,Heinemeyer:2010xt}, and is given by 
%
:
\begin{align}
W &= \sum_{i=1}^{3}\,[~\frac{1}{2}g_{i}^{u}
\,{\bf 10}_i{\bf 10}_i H_{i}+
g_{i}^{d}\,{\bf 10}_i \overline{\bf 5}_{i}\,
\overline{H}_{i}~] +
g_{23}^{u}\,{\bf 10}_2{\bf 10}_3 H_{4} \\
 &+g_{23}^{d}\,{\bf 10}_2 \overline{\bf 5}_{3}\,
\overline{H}_{4}+
g_{32}^{d}\,{\bf 10}_3 \overline{\bf 5}_{2}\,
\overline{H}_{4}+
g_{2}^{f}\,H_{2}\,
{\bf 24}\,\overline{H}_{2}+ g_{3}^{f}\,H_{3}\,
{\bf 24}\,\overline{H}_{3}+
\frac{g^{\lambda}}{3}\,({\bf 24})^3~,\nonumber
\label{w-futb}
\end{align}
while the  solutions to the reduction equations, which ensure the~vanishing of~$\gamma^{(1)}_{i}$, and are \textit{non-degenerate} and~\textit{isolated}~as:
\begin{equation}
\label{zoup-SOL52}
\begin{split}
& (g_{1}^{u})^2
=\frac{8}{5}~ g^2~, ~(g_{1}^{d})^2
=\frac{6}{5}~g^2~,~
(g_{2}^{u})^2=(g_{3}^{u})^2=\frac{4}{5}~g^2~,\\
& (g_{2}^{d})^2 = (g_{3}^{d})^2=\frac{3}{5}~g^2~,~
(g_{23}^{u})^2 =\frac{4}{5}~g^2~,~
(g_{23}^{d})^2=(g_{32}^{d})^2=\frac{3}{5}~g^2~,\\
& (g^{\lambda})^2 =\frac{15}{7}g^2~,~ (g_{2}^{f})^2
=(g_{3}^{f})^2=\frac{1}{2}~g^2~,~ (g_{1}^{f})^2=0~,~
(g_{4}^{f})^2=0~.
\end{split}
\end{equation}
Regarding the SSB sector of the model, assuming the existence of a RGI surface on which Eq.~(\ref{h2NEW}) holds, we obtain at one-loop the generic 
relation~$h=-MC$,~while~the sum rule leads to:
\beq
m^{2}_{H_u}+
2 m^{2}_{{\bf 10}} =M^2~,~
m^{2}_{H_d}-2m^{2}_{{\bf 10}}=-\frac{M^2}{3}~,~
m^{2}_{\overline{{\bf 5}}}+
3m^{2}_{{\bf 10}}=\frac{4M^2}{3}~.
\label{sumrB}
\eeq
As a result there exist two free parameters in the dimensionful sector, $m_{{\bf 10}}$~and $M$.~

After the $SU(5)$ breaking, it is required that the resulting model is the MSSM.  To achieve this, it is necessary to perform a rotation of the Higgs sector, so that the MSSM Higgs doublets are mostly composed from the $5$ and~$\bar 5$ that~couple to~the third~generation.  At the same time, the usual doublet-triplet mechanism has to be implemented to ensure there is no fast proton decay \cite{Leon:1985jm,Kapetanakis:1992vx,Mondragon:1993tw,Hamidi:1984gd, Jones:1984qd,Babu:2002in}. The solutions to the vanishing of the anomalous dimensions (\ref{zoup-SOL52}) and the sum rule (\ref{sumrB}) for the third generation are thus boundary conditions for the MSSM at the GUT scale. The other two generations are minimally coupled to the MSSM Higgs doublets and are therefore taken to zero in this analysis. 
The model is discussed in~more detail in~\cite{Kapetanakis:1992vx,Kubo:1994bj,Mondragon:1993tw,Heinemeyer:2010xt}.

\subsection{Phenomenological Constraints}\label{constraints}

Before the analysis of the above-mentioned model, we will review the experimental constraints applied.\footnote{The~used values do not correspond to the latest~experimental results, which, however, has a~negligible~impact on our analysis.}

We~have consider the pole mass of the top quark while the bottom quark mass~is evaluated at~the $M_Z$ scale, in order to avoid pole mass uncertainties. The experimental values \cite{Tanabashi:2018oca} are:
\beq
m_t^{\rm exp} = 173.1 \pm 0.9 \gev ~~~~~~,~~~~~~ m_b(M_Z) = 2.83 \pm 0.10 \gev~.
\label{mtmbexp}
\eeq
The~Higgs-like particle~discovered in July 2012 by ATLAS and CMS~\cite{Aad:2012tfa,Chatrchyan:2012ufa} is interpreted 
as the light~CP-even Higgs boson~of the MSSM \cite{Heinemeyer:2011aa,Bechtle:2012jw,Bechtle:2016kui}. Its experimental average mass is~\cite{Tanabashi:2018oca}
\beq
M_h^{\rm exp}=125.10\pm 0.14~{\rm GeV}~.\label{higgsexpval}
\eeq
However, it is the theoretical uncertainty \cite{Degrassi:2002fi,Bahl:2017aev} that dominates the total uncertainty, since it is much larger~than the experimental one. For the prediction of the Higgs mass we used the~version 2.16.0 of the~{\tt FeynHiggs} code~\cite{Degrassi:2002fi,Bahl:2017aev,Heinemeyer:1998yj,Heinemeyer:1998np,Frank:2006yh,Hahn:2009zz,Hahn:2013ria,Bahl:2016brp,Bahl:2018qog}.
This version gives a ${\cal O}(2~{\rm GeV})$ downward~shift on the Higgs~mass $M_h$
(for large SUSY masses). More importantly, it gives a reliable point-by-point evaluation~of the uncertainty \cite{Bahl:2019hmm}. The theoretical~uncertainty calculated is~added linearly to the~experimental error of Eq. (\ref{higgsexpval}).

Furthermore, recent ATLAS experiment results  \cite{Aad:2020zxo} limit the neutral Higgs boson masses with respect to $\tan{\beta}$. For our case $\tan{\beta}\sim 45-55$ the lowest limit for the physical neutral Higgs masses is 
\beq M_{A,H}\gtrsim  1900 {\rm ~GeV}.\eeq
For the calculation of the heavy Higgs~sector and the full supersymmetric spectrum a
\texttt{SARAH} \cite{Staub:2013tta} generated, custom module for \texttt{SPheno} \cite{Porod:2003um,Porod:2011nf} was used.~The cross sections for their~particle productions at the HL-LHC~and FCC-hh were calculated~with \texttt{MadGraph5\_aMC@NLO}~\cite{Alwall:2014hca}.

We also~considered the following flavour observables. For $\br(b \to s \gamma)$ we take a value from \cite{Misiak:2006zs,Ciuchini:1998xy,Degrassi:2000qf,Carena:2000uj,DAmbrosio:2002vsn,Asner:2010qj}, while for$\br(B_s \to \mu^+ \mu^-)$ we use a combination~of \cite{Bobeth:2013uxa,Isidori:2012ts,Buras:2003td,Aaij:2012nna,Chatrchyan:2013bka,CMSandLHCbCollaborations:2013pla}:
\beq
\frac{\br(b \to s \gamma )^{\rm exp}}{\br(b \to s \gamma )^{\rm SM}} = 1.089 \pm 0.27~~~~~,~~~~~
\br(B_s \to \mu^+ \mu^-) = (2.9\pm1.4) \times 10^{-9}~.
\eeq
For the~$B_u$ decay to~$\tau\nu$ we use \cite{Isidori:2006pk,Isidori:2007jw,Asner:2010qj,Agashe:2014kda} and~for $\Delta M_{B_s}$ ~we~use \cite{Buras:2000qz,Aaij:2013mpa}:
\beq
\frac{\br(B_u\to\tau\nu)^{\rm exp}}{\br(B_u\to\tau\nu)^{\rm SM}}=1.39\pm 0.69~~~~~~~~~,~~~~~~~~~
\frac{\Delta M_{B_s}^{\rm exp}}{\Delta M_{B_s}^{\rm SM}}=0.97\pm 0.2~.
\eeq

Finally, we~consider Cold Dark Matter (CDM) relic density constraints.~Since the Lightest~SUSY Particle (LSP), which in our case is~the lightest~neutralino, could be a 
promising CDM~candidate \cite{Goldberg:1983nd,Ellis:1983ew}, we examine if~the model is within~the CDM relic density experimental~limits.
The current~bound on the CDM relic density at
$2\,\sigma$ level is given by \cite{Aghanim:2018eyx}%
\beq
\Omega_{\rm CDM} h^2 = 0.1120 \pm 0.0112~.
\label{cdmexp}
\eeq
For the calculation of~the CDM relic density~the {\tt MicrOMEGAs 5.0} code \cite{Belanger:2001fz,Belanger:2004yn,Barducci:2016pcb} was used.

\subsection{Numerical Analysis of the Finite \texorpdfstring{$SU(5)$}{Lg}}\label{analysis}

\begin{figure}[H]
\centering
\includegraphics[width=0.495\textwidth]{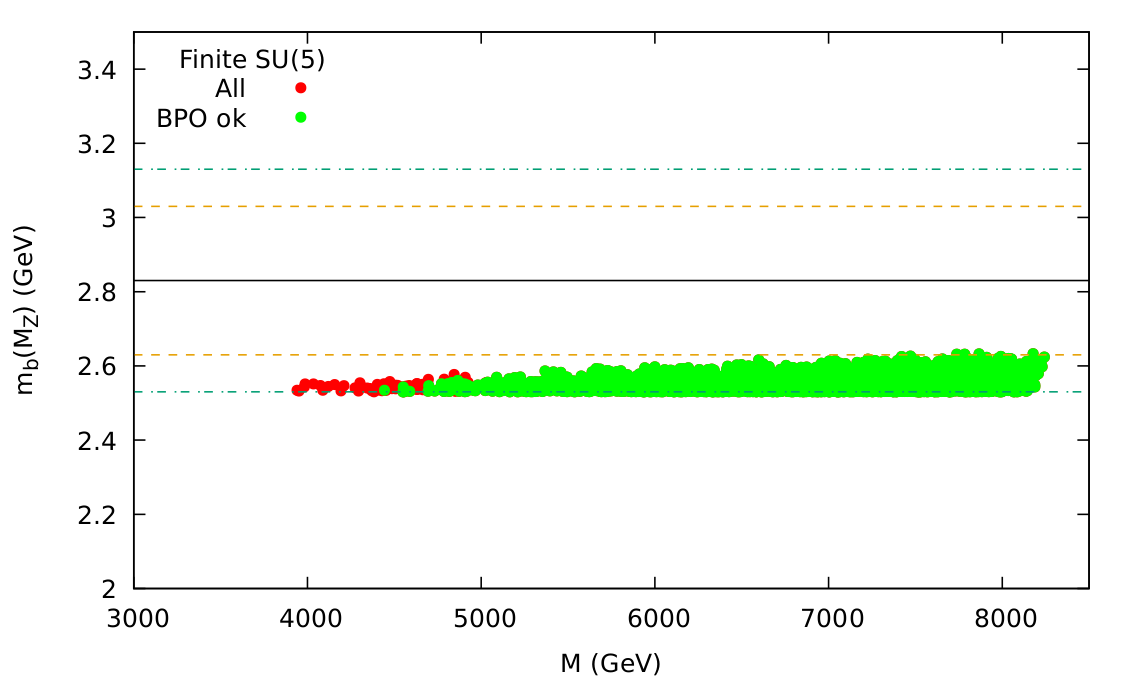}
\includegraphics[width=0.495\textwidth]{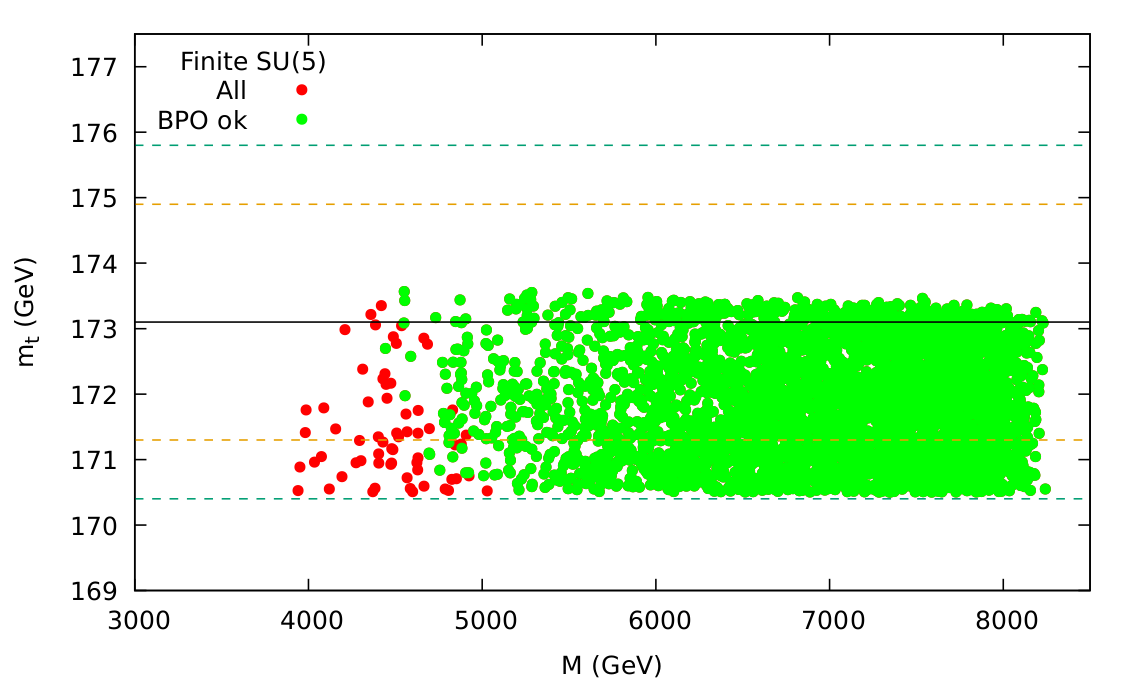}
\caption{\textit{$\mb (M_Z)$~({left})~and $\mt$ ({right}) as a function~of $M$ for the Finite $N=1$ $SU(5)$. The green points~are the ones that satisfy the B-physics constraints.
The~orange (blue) dashed lines denote the 2$\sigma$ (3$\sigma$)~experimental uncertainties.}}
\label{fig:futtopbotvsM}
\end{figure}

We continue with the analysis of~the predicted spectrum of the model.~Below~the GUT scale~we get~the MSSM, where the third generation~is given by the~finiteness conditions (the first two~remain~unrestricted).~However, these conditions do~not restrict~the low-energy~renormalization
properties, so~the above~relations between gauge, Yukawa and the various~dimensionful~parameters serve as boundary conditions at~$M_{GUT}$.~The third generation quark masses
$m_b (M_Z)$~and~$m_t$ are~predicted within 3$\sigma$ and
  2$\sigma$~uncertainties,~respectively, of their experimental values~(see~the complete analysis in \cite{Heinemeyer:2020ftk,Craiova21}),~as shown in \reffi{fig:futtopbotvsM}. 
  The tau lepton mass is used as an input.
  $\mu$ turns out to be negative, as~shown in \cite{Heinemeyer:2012yj,Heinemeyer:2012ai,Heinemeyer:2013fga,Heinemeyer:2018zpw,Heinemeyer:2018bab,
Heinemeyer:2019vbc,Heinemeyer:2019nzo,Heinemeyer:2020ftk,Heinemeyer:2020adl}.

\begin{figure}[htb!]
\centering
\includegraphics[width=0.495\textwidth]{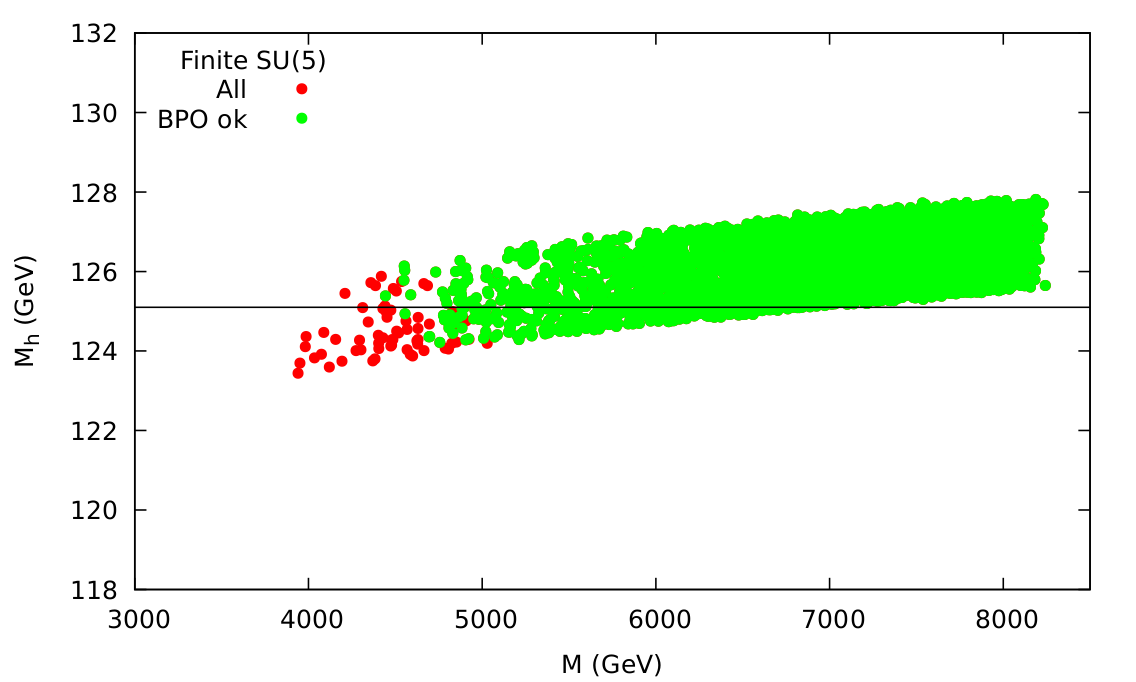}
\includegraphics[width=0.495\textwidth]{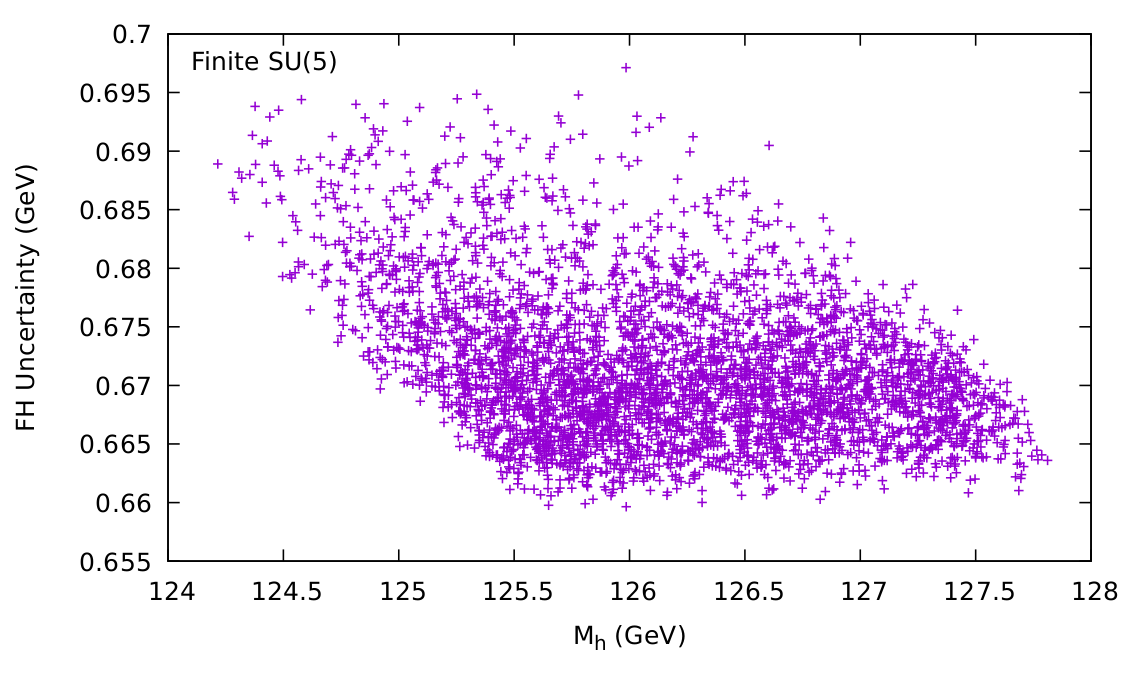}
\caption{\textit{Left: $M_h$~as a function of $M$. Green points comply with~$B$-physics constraints.~Right: The lightest Higgs mass~theoretical uncertainty~calculated with {\tt FeynHiggs}~2.16.0 \cite{Bahl:2019hmm}.}}
\label{fig:futhiggsvsM}
\end{figure}

The plot of~the light Higgs mass satisfies all~experimental constraints~considered in \ref{constraints} (including B-physics~constraints) for a unified gaugino mass $M\sim4500-7500$~GeV, while its point-by-point~theoretical uncertainty \cite{Bahl:2019hmm} drops~significantly (w.r.t. the previous analysis) to~$0.65-0.70$ GeV. This can be found in \reffi{fig:futhiggsvsM}.~The improved~evaluation~of $\Mh$ and its uncertainty prefer~a heavier (Higgs)~spectrum (compared to previous~analyses \cite{Heinemeyer:2010xt,Heinemeyer:2015dpa,Heinemeyer:2012yj,Heinemeyer:2013nza,Heinemeyer:2012ai,Heinemeyer:2013fga,Heinemeyer:2018roq,
Heinemeyer:2018zpw,Heinemeyer:2018bab,Heinemeyer:2019vbc,Heinemeyer:2019nzo,Heinemeyer:2020ftk,Mondragon:2011zzb}), and thus~allows only a heavy~supersymmetric~spectrum, which is in~agreement with~all existing experimental data. Very~heavy coloured~supersymmetric particles are favoured, in agreement with~the non-observation of such particles at the LHC~\cite{2018:59-1,2018:59-2}.

Concerning CDM, although no point fulfills the strict~bound~of \refeq{cdmexp}, since we have overproduction of~CDM in the early universe (for~the original analysis see~\cite{Patellis:2021drd}), we can extend the model by considering~bilinear~R-parity
violating~terms (that preserve finiteness) and thus introduce neutrino masses~\cite{Valle:1998bs,Diaz:2003as}.
R-parity violation \cite{Dreiner:1997uz,Bhattacharyya:1997vv,Allanach:1999ic,Romao:1991ex}~would remove~the CDM~bound~of
\refeq{cdmexp} completely.

As explained~in more~detail in \cite{Heinemeyer:2020nzi}, the three~benchmarks chosen~(for the purposes of collider phenomenology)~feature the LSP above
$2100$~GeV,~$2400$~GeV~and $2900$~GeV, respectively.
The resulting~masses~that are relevant to our~analysis were generated by~{\tt SPheno} 4.0.4
\cite{Porod:2003um,Porod:2011nf}~and are listed in \refta{tab:futbspheno}~for~each benchmark (with the corresponding~$\tan\beta$).~The two first masses refer to the heavy Higgs~bosons. The~gluino mass is $M_{\tilde{g}}$, the~neutralinos and~the~charginos are~denoted as $M_{\tilde{\chi}_i^0}$~and~$M_{\tilde{\chi}_i^{\pm}}$, while~the slepton~and sneutrino masses~for all three~generations~are given~as
$M_{\tilde{e}_{1,2,3}},~M_{\tilde{\nu}_{1,2,3}}$. Similarly,~the~squarks
are denoted~as $M_{\tilde{d}_{1,2}}$~and $M_{\tilde{u}_{1,2}}$~for the~first two generations. The third~generation~masses~are given by
$M_{\tilde{t}_{1,2}}$~for stops~and~$M_{\tilde{b}_{1,2}}$ for sbottoms.

\begin{center}
\begin{table}[ht]
\begin{center}
\small
\begin{tabular}{|l|r|r|r|r|r|r|r|r|r|r|r|r|}
\hline
  & $tan\beta$ & $M_{A,H}$ & $M_{H^{\pm}}$  & $M_{\tilde{g}}$ & $M_{\tilde{\chi}^0_1}$ & $M_{\tilde{\chi}^0_2}$ & $M_{\tilde{\chi}^0_3}$  & $M_{\tilde{\chi}^0_4}$ &  $M_{\tilde{\chi}_1^\pm}$ & $M_{\tilde{\chi}_2^\pm}$  \\\hline
FUTSU5-1 & 49.9 & 5.688 & 5.688  & 8.966 & 2.103 & 3.917 & 4.829 & 4.832 & 3.917 & 4.833  \\\hline
FUTSU5-2 & 50.1 & 7.039 &  7.086 & 10.380 & 2.476 & 4.592 & 5.515 & 5.518 & 4.592 & 5.519  \\\hline
FUTSU5-3 & 49.9 & 16.382 &  16.401 & 12.210 & 2.972 & 5.484 & 6.688 & 6.691 & 5.484 & 6.691  \\\hline
 & $M_{\tilde{e}_{1,2}}$ & $M_{\tilde{\nu}_{1,2}}$ & $M_{\tilde{\tau}}$ & $M_{\tilde{\nu}_{\tau}}$ & $M_{\tilde{d}_{1,2}}$ & $M_{\tilde{u}_{1,2}}$ & $M_{\tilde{b}_{1}}$ & $M_{\tilde{b}_{2}}$ & $M_{\tilde{t}_{1}}$ & $M_{\tilde{t}_{2}}$ \\\hline
FUTSU5-1 & 3.102 & 3.907 & 2.205 & 3.137 & 7.839 & 7.888 & 6.102 & 6.817 & 6.099 & 6.821 \\\hline
FUTSU5-2 & 3.623 & 4.566 & 2.517 & 3.768 & 9.059 & 9.119 & 7.113 & 7.877 & 7.032 & 7.881 \\\hline
FUTSU5-3 & 4.334 & 5.418 & 3.426 & 3.834 & 10.635 & 10.699 & 8.000 & 9.387 & 8.401 & 9.390 \\\hline
\end{tabular}
\caption{Masses~for~each of~the three~benchmarks of the Finite $N=1$ $SU(5)$ (in~TeV)~\cite{Heinemeyer:2020nzi}.}\label{tab:futbspheno}
\end{center}
\end{table}
\end{center}

At 14 TeV~HL-LHC none~of the  Finite $SU(5)$ scenarios listed above has a~SUSY~production cross section above 0.01~fb, and thus will~most~probably remain~unobservable~\cite{Cepeda:2019klc}.~The~discovery prospects for~the heavy Higgs-boson spectrum~is significantly better at the~FCC-hh \cite{Hajer:2015gka}.~Theoretical analyses \cite{Craig:2016ygr,Hajer:2015gka} have~shown that for large~$\tb$ heavy~Higgs mass~scales up to $\sim 8$ TeV could be~accessible.~Since in this model we have $\tb \sim 50$, the~first two~benchmark points are well within the reach of~the~FCC-hh (as explained in \cite{Heinemeyer:2020nzi}). The third~point, however, where $\MA \sim 16$~TeV,~will be far outside~the reach~of the collider.
At 100~TeV we have in principle production of SUSY particles in
pairs, although their
production~cross~section is~at the few fb~level .  This is a result of~the heavy~spectrum~of the model. Comparing our benchmark~predictions with the~simplified model limits of \cite{Golling:2016gvc},~we have~found that the lighter stop might be~accessible in~FUTSU5-1 (see \cite{Heinemeyer:2020nzi}).~For~the squarks of the~first two generations~there are better prospects.~All benchmarks~could be~tested at~the $2\,\sig$ level,~but no discovery at~the $5\,\sig$ can~be expected and the~same holds for the~gluino.

\section{Conclusions}\label{conclusions}

Veltman's contributions to the field of Particle Physics have a huge impact in the development of the field. Here we have presented only one of the roads that Veltman opened in particle physics, published in his celebrated paper in Acta Physica Polonica \cite{Veltman:1980mj}. This work of Veltman was guided by the current at that time notion of naturality which appeared to be of fundamental importance in physics discussions and guiding principle in searches of new physics up to now. Veltman required the absence of quadratic divergences in the SM, which led to a quadratic mass relation among the SM particles. 
The fact that Veltman’s relation eventually does not hold can be
taken as sign that the SM, despite its phenomenological successes
cannot be considered as a complete theory.
 Moreover, the whole discussion on the cancellation of quadratic divergences in renormalizable field theories with scalars was uniquelly pointing to the supersymmetric ones, where naturally do not appear such divergences to all-orders of perturbation theory, as the arena of searching for a more complete theory. The possibility to achieve unification of the gauge couplings of the SM in the supersymmetric framework with supersymmetry broken in the TeV scale gave a huge push in the research in such theories and in particular in MSSM for many years.

Still, the problem of the several independent parameters of the SM is much more severe even in the minimal version of MSSM, when the supersymmetry breaking sector is taken into account. Correspondingly the necessity of \textit{reduction of couplings} in the SM becomes substantially stronger in the MSSM. The application of the method of searching for RGI relations as a way to reduce the independent parameters of the SM failed, as the Veltman relation, when it was confronted with the experimental discoveries of the top quark and Higgs particles and the determination of their masses. Now after several years of research it seems that so far supersymmetric unified schemes such as the Finite $SU(5)$, the minimal $SU(5)$ and the $SU(3)^3$ with reduced couplings (i.e. satifying RGI relations) can be realistic. Among them clearly the most interesting is the $SU(5)$ FUT, since beyond the unification scale  a complete reduction of couplings in favour of the gauge coupling can be achieved, and it is furthermore finite to all-orders in perturbation theory. In the latter clearly even the logarithmic divergences are absent, fulfilling an old dream of theoretical physicists who were seriously disturbed by the presence of divergences in field theories. Moreover, it is a realistic theory with the great successes of predicting successfully the top and Higgs masses well before their experimental discoveries and passing successfully all experimental tests so far and having chances to be tested further at FCC-hh.

\section*{Acknowledgements}

One of us (GZ) would like to thank from the heart Tini Veltman for his long standing friendship for which he is indeed very honoured. All of us we would like to express our deepest condolences to his family and in particular to his wife Anneke and daughter Helene. Also, we would like to
thank Marek Jeżabek and Michał Praszałowicz for the invitation to
contribute in the present special volume in honour of Martinus J. G.
Veltman.

We would like to thank our collaborators  
Sven Heinemeyer,
Jan Kalinowski,
Dimitris Kapetanakis,
Tatsuo Kobayashi, 
Wojciech Kotlarski,
Jisuke Kubo, 
Ernest Ma,
Nick Tracas, for the long standing collaboration, 
and Wolfgang Hollik and Dieter L\"{u}st, for their constant encouragement in the
Finiteness project.

This work has been supported by the Basic Research Programme, PEVE2020 of the National Technical University of Athens, Greece.
It was~partially supported~by DGAPA grant PAPIIT IN109321. The work~of GP and~GZ is partially~supported~by~the COST~action~CA16201, GZ is also partially~supported by~the~grant DEC-2018/31/B/ST2/02283 of NSC, Poland. GZ would like to thank the  DFG Exzellenzcluster 2181:STRUCTURES of Heidelberg University, MPI, CERN-TH and the Excellence Grant~Enigmass~of LAPTh for support.

\bibliographystyle{h-physrev5}
\bibliography{biblio-RoC}
\end{document}